\documentclass[twocolumn]{jpsj3}
\usepackage{graphicx}
\usepackage{color}
\usepackage{bm}

\title{%
Boundary Potential Method for Describing Electron Teleportation
in an Interferometer with a Topological Superconductor
}

\author{%
Kyosuke Mizuno, Yuto Takarabe, and Yositake Takane$^{*}$
}

\inst{%
Graduate School of Advanced Science and Engineering,\\
Hiroshima University, Higashihiroshima, Hiroshima 739-8530, Japan
}

\recdate{ \hspace{50mm} }

\abst{%
One-dimensional topological superconductors accommodate
a pair of Majorana zero modes at their ends.
In an interferometer containing such a topological superconductor,
electron transport is significantly affected by the Majorana zero modes
constituting a nonlocal state localized near both ends of the superconductor.
When the number of electrons $\mathcal{N}$ in the superconductor is
constrained by a charging effect, the resonant tunneling through
the nonlocal state is expected to result in unusual transport properties.
This resonant tunneling, called electron teleportation, is not easy to describe
because there is no simple method to handle the constraint on $\mathcal{N}$.
Here, we propose a boundary potential method based on scattering theory
for calculating the conductance of the interferometer
under a given constraint on $\mathcal{N}$.
This method enables us to calculate the conductance
taking account of relevant charging energy and details of the system.
}

\begin{document}
%%\sloppy
\maketitle

\section{Introduction}

Topological superconductors accommodate Majorana zero modes
at their boundary.~\cite{Kitaev,Alicea,Sato1}
A Majorana zero mode is an equal superposition of
electron and hole states appearing at zero energy.
Its creation and annihilation operators are identical, indicating that
the degree of freedom of a Majorana zero mode is less than
that of a single fermion.
Indeed, two Majorana zero modes are necessary to constitute
the degree of freedom of a single fermion.
One of the target systems for studies on a Majorana zero mode
is a one-dimensional topological superconductor constituted
by a semiconductor nanowire with strong spin-orbit interaction
and a Zeeman field under the proximity effect
from $s$-wave superconductivity.~\cite{Lutchyn,Oreg,Mourik}
A Majorana zero mode at zero energy is localized near each end of
this topological superconductor.
Theoretical and experimental attention has been paid to electron transport
in this type of proximity-induced topological superconductor,
each end of which is connected to a normal metal lead.
The two Majorana zero modes are hybridized to form a nonlocal state
localized near the two ends of the superconductor
with energy nearly equal to zero.~\cite{Nilsson,fu1,Akhmerov}
We expect that this nonlocal state induces the resonant tunneling of
an electron through the superconductor.
However, this resonant tunneling, referred to as electron teleportation,
is difficult to detect in an ordinary setup
because an electron incident into the superconductor from a normal metal lead
is reflected as a hole with a probability of nearly one.
That is, Andreev reflection from an electron to a hole occurs nearly perfectly
in the presence of a Majorana zero mode at the interface.

%%%%%%%%%%%%%%%%%%
\begin{figure}[btp]
\begin{center}
\includegraphics[height=1.5cm]{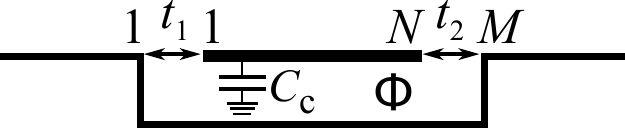}
\end{center}
\caption{
Interferometer consisting of a topological superconductor wire of $N$ sites,
which is connected to ground by a capacitor,
and an infinitely long normal metal lead.
The left (i.e., 1st site) and right (i.e., $N$th site) ends
of the superconductor are
connected to the 1st and $M$th sites of the normal metal lead, respectively,
where $t_{1}$ ($t_{2}$) represents the transfer integral between the $1$st
($N$th) site in the superconductor and the $1$st ($M$th) site
in the normal metal lead.
A magnetic flux $\Phi$ pierces the loop formed by
the superconductor and the normal metal lead.
}
\end{figure}
%%%%%%%%%%%%%%%%%%
This situation changes if the topological superconductor is
of a mesoscopic size and is grounded by a capacitor of capacitance $C_{\rm c}$.
The charging energy of the superconductor depends on the number of electrons
$\mathcal{N}$ in it as
\begin{align}
  U(\mathcal{N})=\frac{1}{2C_{\rm c}}\left(\mathcal{N}e-Q_{0}\right)^{2} ,
\end{align}
where $Q_{0}$ is the gate charge
determined by the voltage across the capacitor.
We assume that the typical charging energy $\frac{e^{2}}{2C_{\rm c}}$ is
nearly equal to the energy gap of the superconductor $E_{\rm g}$ and that
the temperature $T$ satisfies $k_{\rm B}T \ll E_{g}$,
where $k_{\rm B}$ is the Boltzmann constant.
Under the charging effect, $\mathcal{N}$ in the ground state of
the superconductor is not necessarily even
since the Majorana zero modes are present.
The nonlocal state consisting of the Majorana zero modes is empty
in the ground state with $\mathcal{N} = 2l$ and is occupied in the ground state
with $\mathcal{N} = 2l+1$, where $l$ is an integer.
Let us assume $Q_{0}/e \approx 2l+\frac{1}{2}$.
Under this assumption,
$\mathcal{N}$ is constrained to vary between $2l$ and $2l+1$.
This constraint prohibits Andreev reflection processes,
which increase or decrease $\mathcal{N}$ by two.
Without these processes, the electron teleportation can affect
the transport properties of electrons through the superconductor.~\cite{fu2}

For the experimental detection of
the electron teleportation,~\cite{fu2,Vijay,Hell,
Whiticar,Sugeta,Goto} an interferometer consisting of a topological
superconductor and a normal metal lead is a suitable setup (see Fig.~1).
This interferometer is pierced by a flux $\Phi$,
and $\mathcal{N}$ in the superconductor is constrained by the charging effect.
As a result of the electron teleportation,
the zero-bias conductance of the interferometer
oscillates as a function of $\Phi$ with a period of $\Phi_{0}=h/e$
and the phase of this oscillation shifts by $\pi$ when $\mathcal{N}$
in the ground state changes from $2l$ to $2l\pm1$.~\cite{fu2}
This feature serves as one evidence for the presence of
a pair of Majorana zero modes.
If the constraint due to the charging effect is absent,
the zero-bias conductance oscillates as a function of $\Phi$ with a period of
$\Phi_{0}/2=h/2e$ as a result of Andreev reflection processes.~\cite{takane1}
This oscillation arises even if the topological superconductor is replaced with an ordinary superconductor.

The unusual electron transport due to the electron teleportation provides
useful information on the Majorana zero modes.
However, it is not easy to describe the electron teleportation theoretically
because there is no simple method for handling the constraint on $\mathcal{N}$.
Although various techniques have been used to address
this problem,~\cite{Zazunov,Hutzen,Heck,Chiu,Ekstrom,Jin,Shapiro,Kleinherbers}
they cannot be simply adapted to the interferometer
or require time-consuming computations if possible.
Goto {\it et. al.}~\cite{Goto} calculated the zero-bias
conductance of the interferometer without using such techniques.
Ignoring the charging effect, they determined the excitation spectrum
of the isolated loop structure in the interferometer as a function of $\Phi$
and classified the eigenvalues of energy into two groups,
one of which is consistent with the ground state with $\mathcal{N} = 2l$
and the other is consistent with the ground state with $\mathcal{N} = 2l\pm 1$.
The zero-bias conductance in the ground state
with $\mathcal{N} = 2l$ ($2l\pm 1$) is calculated
using the eigenvalues in the first (second) group.
This method is notable as a practical one applicable to the interferometer.
However, it does not directly control the variation of $\mathcal{N}$
and cannot incorporate relevant charging energy.

Let us consider the charging effect in more detail.
Hereafter, we denote the ground and excited states with $\mathcal{N}$
as ${\rm G}_{\mathcal{N}}$ and ${\rm E}_{\mathcal{N}}$, respectively.
To specify $Q_{0}$,
we use $Q_{2l+\frac{\xi}{2}}$, which is defined with an integer $\xi$ as
\begin{align}
  Q_{2l+\frac{\xi}{2}} = \left(2l+\frac{\xi}{2}\right)e ,
\end{align}
and a positive value $\delta Q$ satisfying $0 < \delta Q/e \ll \frac{1}{2}$.
If the gate charge is $Q_{0} = Q_{2l+\frac{1}{2}} -\delta Q$,
the ground state ${\rm G}_{2l}$ is nearly degenerate with
the first excited state ${\rm E}_{2l+1}$ and
\begin{align}
        \label{eq:def-dU+}
 \delta U \equiv U(2l+1) - U(2l)
          = \frac{e \delta Q}{C_{\rm c}} > 0.
\end{align}
In this case, $\mathcal{N}$ is constrained to vary between $2l$ and $2l+1$
in the scattering processes of a quasiparticle.
We express this constraint as ${\rm G}_{2l} \leftrightarrow {\rm E}_{2l+1}$.
If $Q_{0} = Q_{2l+\frac{1}{2}} + \delta Q$,
the ground state ${\rm G}_{2l+1}$ is nearly degenerate with
the first excited state ${\rm E}_{2l}$, and the constraint changes to
${\rm G}_{2l+1} \leftrightarrow {\rm E}_{2l}$.
The ground state is not uniquely determined in the limit of $\delta U \to 0$.
If the gate charge is $Q_{0} = Q_{2l-\frac{1}{2}} + \delta Q$,
the ground state ${\rm G}_{2l}$ is nearly degenerate with
the first excited state ${\rm E}_{2l-1}$ and
\begin{align}
        \label{eq:def-dU-}
 \delta U \equiv U(2l-1) - U(2l)
          = \frac{e \delta Q}{C_{\rm c}} > 0.
\end{align}
The constraint is expressed as ${\rm G}_{2l} \leftrightarrow {\rm E}_{2l-1}$.
If $Q_{0} = Q_{2l-\frac{1}{2}} - \delta Q$,
the ground state ${\rm G}_{2l-1}$ is nearly degenerate with
the excited state ${\rm E}_{2l}$,
and the constraint changes to ${\rm G}_{2l-1} \leftrightarrow {\rm E}_{2l}$.
Again, the ground state is not uniquely determined
in the limit of $\delta U \to 0$.

The above consideration shows that $\delta U$ is a significant parameter
that characterises the charging effect.
Therefore, it is important to elucidate the effect of $\delta U$
on the conductance of the interferometer.
In this paper, we propose a boundary potential method
based on scattering theory for calculating the conductance
of the interferometer under a given constraint on $\mathcal{N}$.
This method directly controls the variation of $\mathcal{N}$, enabling us
to incorporate the effects of $\delta U$ and the details of the system.

In the next section, we introduce the Hamiltonian of the interferometer
consisting of a one-dimensional proximity-induced topological superconductor
and a normal metal lead.
This interferometer is regarded as a piecewise superconducting loop
attached to the left and right normal metal leads.
In Sect.~3, we obtain the boundary potential that fully describes
the effects of the superconductor on the normal metal lead
in the absence of the charging effect
and formulate the scattering problem of an electron
incident into the interferometer.
In Sect.~4, the charging effect on the scattering problem is considered.
We modify the boundary potential obtained in Sect.~3 such that it properly
describes the constraint on $\mathcal{N}$ including the effect of $\delta U$.
In Sect.~5, we give numerical results on the nonlocal conductance
of the interferometer at zero temperature.
The last section is devoted to a summary and discussion.

\section{Model and Symmetry}

We consider an interferometer consisting of
a one-dimensional topological superconductor,
which accommodates a pair of Majorana zero modes at its ends,
and a normal metal lead (see Fig.~1).
We express the Hamiltonian for this interferometer as
\begin{align}
  H = H_{\rm TS} + H_{\rm N} + H_{\rm T} ,
\end{align}
where $H_{\rm TS}$ and $H_{\rm N}$ describe the topological superconductor
and the normal metal lead, respectively, each of which is defined on
a one-dimensional lattice with lattice constant $a$, and $H_{\rm T}$ represents
the coupling between the topological superconductor and the normal metal lead.
On a one-dimensional doubly connected lattice structure,
we use $i$ and $j$ to specify the sites in the topological superconductor
and the normal metal lead, respectively.
We use $c_{i\sigma}^{\dagger}$ and $c_{i\sigma}$  to represent the creation
and annihilation operators of the electron with spin $\sigma$
at the $i$th site in the topological superconductor
and $f_{j\sigma}^{\dagger}$ and $f_{j\sigma}$ to represent the creation and
annihilation operators of the electron with spin $\sigma$
at the $j$th site in the normal metal lead, respectively.

The Hamiltonian for the topological superconductor
is given by~\cite{Lutchyn,Oreg,Sato2}
\begin{align}
     \label{eq:def-H_TS}
  H_{\rm TS}
 & = \sum_{i=1}^{N-1}\sum_{\sigma=\uparrow,\downarrow}
     \left( -t c_{i+1\sigma}^{\dagger}c_{i\sigma}
            -t c_{i\sigma}^{\dagger}c_{i+1\sigma} \right)
       \nonumber \\
 & \hspace{-6mm}
   + \sum_{i=1}^{N-1}
     \left(  \lambda c_{i+1\downarrow}^{\dagger}c_{i\uparrow}
            +\lambda c_{i\uparrow}^{\dagger}c_{i+1\downarrow}
            -\lambda c_{i\downarrow}^{\dagger}c_{i+1\uparrow}
            -\lambda c_{i+1\uparrow}^{\dagger}c_{i\downarrow} \right)
       \nonumber \\
 & \hspace{-6mm}
   + \sum_{i=1}^{N}
     \left(  \Delta c_{i\uparrow}^{\dagger}c_{i\downarrow}^{\dagger}
            +\Delta c_{i\downarrow}c_{i\uparrow}
            +h c_{i\uparrow}^{\dagger}c_{i\uparrow}
            -h c_{i\downarrow}^{\dagger}c_{i\downarrow} \right)
       \nonumber \\
 & \hspace{-6mm}
   + \sum_{i=1}^{N}\sum_{\sigma=\uparrow,\downarrow}
     \left(-\mu_{\rm S} c_{i\sigma}^{\dagger}c_{i\sigma}\right) ,
\end{align}
where $t$ is the transfer integral between  nearest-neighboring sites,
and $\lambda$, $h$, $\Delta$, and $\mu_{\rm S}$ are respectively
the strength of the spin-orbit interaction, the strength of the Zeeman field,
the pair potential describing the proximity effect,
and the chemical potential in the superconductor.~\cite{Sugeta,Goto}
The Hamiltonian for the normal metal lead is
\begin{align}
     \label{eq:def-H_N}
  H_{\rm N}
   = \sum_{j=-\infty}^{+\infty}\sum_{\sigma=\uparrow,\downarrow}
     \left( -t' f_{j+1\sigma}^{\dagger}f_{j\sigma}
            -t' f_{j\sigma}^{\dagger}f_{j+1\sigma}
            -\mu_{\rm N} f_{j\sigma}^{\dagger}f_{j\sigma} \right) ,
\end{align}
where $t'$ is the transfer integral between nearest-neighboring sites
and $\mu_{\rm N}$ is the chemical potential in the normal metal lead.
The coupling between the topological superconductor and the normal metal lead
is described by
\begin{align}
     \label{eq:def-H_T}
  H_{\rm T}
 & = \sum_{\sigma=\uparrow,\downarrow}
     \left( -t_{1} c_{1\sigma}^{\dagger}f_{1\sigma}
            -t_{1} f_{1\sigma}^{\dagger}c_{1\sigma} \right)
       \nonumber \\
 & \hspace{-6mm}
   + \sum_{\sigma=\uparrow,\downarrow}
     \left( -t_{2}e^{-i\phi} c_{N\sigma}^{\dagger}f_{M\sigma}
            -t_{2}e^{i\phi} f_{M\sigma}^{\dagger}c_{N\sigma} \right) ,
\end{align}
where $t_{1}$ ($t_{2}$) represents the transfer integral between
the $1$st ($N$th) site in the topological superconductor
and the $1$st ($M$th) site in the normal metal lead.
We take account of the magnetic flux $\Phi$ piercing the interferometer
by replacing $t_{2}$ with $t_{2}e^{\pm i\phi}$, where the phase $\phi$
is defined as
\begin{align}
  \phi = 2\pi\frac{\Phi}{\Phi_{0}} .
\end{align}

We here consider the spectrum of the topological superconductor
isolated from the normal metal lead, which is described by $H_{\rm TS}$.
In terms of the column and row vectors defined by
\begin{align}
   C
 & = \bigl[c_{1\uparrow}\, c_{1\downarrow}\,
                   c_{1\uparrow}^{\dagger}\, c_{1\downarrow}^{\dagger} \dots
                   c_{N\uparrow}\, c_{N\downarrow}\,
                   c_{N\uparrow}^{\dagger}\, c_{N\downarrow}^{\dagger}
             \bigr]^\mathsf{T} ,
         \\
   C^{\dagger}
 & = \bigl[c_{1\uparrow}^{\dagger}\, c_{1\downarrow}^{\dagger}\,
           c_{1\uparrow}\, c_{1\downarrow} \dots
           c_{N\uparrow}^{\dagger}\, c_{N\downarrow}^{\dagger}\,
           c_{N\uparrow}\, c_{N\downarrow}
     \bigr] ,
\end{align}
we rewrite $H_{\rm TS}$ as $H_{\rm TS} = \frac{1}{2}C^{\dagger}h_{\rm TS}C$
with the $4N \times 4N$ symmetric matrix $h_{\rm TS}$ given by
\begin{align}
     \label{eq:h-TS}
   h_{\rm TS} =
 \left[
   \begin{array}{ccccc}
     h_{0} & h_{1}^\mathsf{T} & 0_{4 \times 4} & 0_{4 \times 4} & \cdots \\
     h_{1} & h_{0} & h_{1}^\mathsf{T} & 0_{4 \times 4} & \cdots \\
     0_{4 \times 4} & h_{1} & h_{0} & h_{1}^\mathsf{T} & \cdots \\
     0_{4 \times 4} & 0_{4 \times 4} & h_{1} & h_{0} & \cdots \\
     \cdots & \cdots & \cdots & \cdots & \cdots
   \end{array}
  \right] ,
\end{align}
where
\begin{align}
  h_{0} & = \left[ \begin{array}{cccc}
                     h-\mu_{\rm S} & 0 & 0 & \Delta \\
                     0 & -h-\mu_{\rm S} & -\Delta & 0 \\
                     0 & -\Delta & -h+\mu_{\rm S} & 0 \\
                     \Delta & 0 & 0 & h+\mu_{\rm S}
                     \end{array}
            \right] ,
     \\
  h_{1} & = \left[ \begin{array}{cccc}
                     -t & -\lambda & 0 & 0 \\
                     \lambda & -t & 0 & 0 \\
                     0 & 0 & t & \lambda \\
                     0 & 0 & -\lambda & t
                   \end{array}
            \right] .
\end{align}
The matrix $h_{\rm TS}$ is regarded as the Hamiltonian
describing a Bogoliubov--de Gennes (BdG) equation for
quasiparticles (i.e., bogolons) in the topological superconductor.
Its particle--hole symmetry is represented as
\begin{align}
  \Xi h_{\rm TS} \Xi = - h_{\rm TS} ,
\end{align}
where $\Xi = 1_{N \times N} \otimes \alpha_{1}$ with
\begin{align}
  \alpha_{1} = \left[ \begin{array}{cc}
                         0_{2 \times 2} & 1_{2 \times 2} \\
                         1_{2 \times 2} & 0_{2 \times 2}
                      \end{array}
               \right] .
\end{align}
This symmetry ensures that
an eigenvector $|\varphi\rangle$ of $h_{\rm TS}$ satisfying
$h_{\rm TS}|\varphi\rangle = E|\varphi\rangle$
is paired with another eigenvector
$|\tilde{\varphi}\rangle = \Xi|\varphi\rangle$ satisfying
$h_{\rm TS} |\tilde{\varphi}\rangle = -E|\tilde{\varphi}\rangle$.
This relation ensures that $4N$ eigenvalues of $h_{\rm TS}$ are written as
$\pm E_{1}$, $\pm E_{2}$, $\dots$, $\pm E_{2N}$
with $0 \le E_{1} \le E_{2} \le \cdots \le E_{2N}$.
Here, $\pm E_{1}$ corresponds to the nonlocal state
comprising a pair of Majorana zero modes.

Let us introduce the eigenvectors defined by
\begin{align}
 h_{\rm TS}|\varphi_{n}\rangle & = E_{n}|\varphi_{n}\rangle ,
     \\
 h_{\rm TS}|\varphi_{-n}\rangle & = -E_{n}|\varphi_{-n}\rangle
\end{align}
for $n = 1,2,\dots, 2N$.
They satisfy $\langle \varphi_{n}|\varphi_{m}\rangle = \delta_{n,m}$
and
\begin{align}
 |\varphi_{-n}\rangle = \Xi|\varphi_{n}\rangle .
\end{align}
For later convenience, we explicitly express
$|\varphi_{n}\rangle$ as
\begin{align}
      \label{eq:elements-varphi}
  |\varphi_{n}\rangle = \left[\begin{array}{c}
                                u_{n\uparrow}(1) \\
                                u_{n\downarrow}(1) \\
                                v_{n\uparrow}(1) \\
                                v_{n\downarrow}(1) \\
                                \vdots \\
                                u_{n\uparrow}(N) \\
                                u_{n\downarrow}(N) \\
                                v_{n\uparrow}(N) \\
                                v_{n\downarrow}(N)
                              \end{array}
                        \right] .
\end{align}

Let us define the $4N \times 4N$ matrices $V$ and $V^{\dagger}$ as
\begin{align}
      \label{eq:V}
   V & = \left[ |\varphi_{1}\rangle \, |\varphi_{-1}\rangle \cdots
              |\varphi_{2N}\rangle \, |\varphi_{-2N}\rangle
         \right] ,
   \\
      \label{eq:V-dag}
   V^{\dagger} & = \left[ \begin{array}{c}
                            \langle\varphi_{1}| \\
                            \langle\varphi_{-1}| \\
                            \vdots \\
                            \langle\varphi_{2N}| \\
                            \langle\varphi_{-2N}| \\
                          \end{array}
                   \right] .
\end{align}
In terms of $V$ and $V^{\dagger}$, $h_{\rm TS}$ is diagonalized as
\begin{align}
  V^{\dagger}h_{\rm TS}V =
  {\rm diag}\left(E_{1}, -E_{1}, \dots, E_{2N},  -E_{2N}\right) .
\end{align}
Let us define the operators describing the bogolons as
\begin{align}
       \label{eq:def-d}
      \left[ d_{1}^{\dagger} \, d_{1} \, d_{2}^{\dagger} \, d_{2}
             \dots d_{2N}^{\dagger} \, d_{2N}
      \right]
 & = C^{\dagger}V ,
          \\
       \label{eq:def-d-dag}
      \left[ d_{1} \, d_{1}^{\dagger} \, d_{2} \, d_{2}^{\dagger}
             \dots d_{2N} \, d_{2N}^{\dagger}
      \right]^\mathsf{T}
 & = V^{\dagger}C .
\end{align}
These operators obey the anticommutation relations:
\begin{align}
 \left\{ d_{n}, d_{m}^{\dagger}\right\} = \delta_{n,m} ,
   \hspace{5mm}
 \left\{ d_{n}, d_{m}\right\}
 = \left\{ d_{n}^{\dagger}, d_{m}^{\dagger}\right\} = 0.
\end{align}
Using these operators, we can express $H_{\rm TS}$ as
\begin{align}
      \label{eq:H-bogolon}
   H_{\rm TS} = \sum_{n=1}^{2N}E_{n}d^{\dagger}_{n}d_{n} ,
\end{align}
where a constant term is subtracted.
Here, $E_{1} > 0$ is implicitly assumed in defining the operators
in Eqs.~(\ref{eq:def-d}) and (\ref{eq:def-d-dag}).
Indeed, if the length $N$ of the topological superconductor is finite,
the two Majorana zero modes at both ends of the topological superconductor
should be hybridized to form the nonlocal state at $E_{1} > 0$.

Here, we comment on a practical procedure for determining
the eigenvectors $|\varphi_{1}\rangle$ and $|\varphi_{-1}\rangle$
for the nonlocal state consisting of the left and right Majorana zero modes.
Let us assume that we obtained two near-zero eigenvalues, $e_{1}$ and $e_{2}$,
and their corresponding eigenvectors, $|\chi_{1}\rangle$ and
$|\chi_{2}\rangle$, through double-precision calculation.
If they satisfy $e_{1} = - e_{2} > 0$
and $|\chi_{2}\rangle = \Xi|\chi_{1}\rangle$ with a numerical accuracy
determined by double-precision calculation,
we can identify $|\varphi_{1}\rangle=|\chi_{1}\rangle$ and
$|\varphi_{-1}\rangle=|\chi_{2}\rangle$
with $E_{1} = e_{1} = - e_{2}$.
This identification cannot be allowed when $N$ is sufficiently large,
so that the hybridization of the two Majorana zero modes becomes
negligibly small, and therefore, the true eigenvalue $E_{1}$
normalized by $t$ is nearly equal to or less than $1.0 \times 10^{-14}$.
This inevitably leads to the problem that $e_{1} \neq - e_{2}$
and $|\chi_{2}\rangle \neq \Xi|\chi_{1}\rangle$, indicating that
the particle--hole symmetry is broken owing to numerical imprecision
near the degeneracy point.
Since $e_{1}$ and $e_{2}$ are extremely small, their values are unreliable.
However, $|\chi_{1}\rangle$ and $|\chi_{2}\rangle$ are reliable
as near-zero eigenvectors.
The relation $|\chi_{2}\rangle = \Xi|\chi_{1}\rangle$ is broken because
$|\chi_{1}\rangle$ and $|\chi_{2}\rangle$ are formed by a certain mixture of
the true eigenvectors $|\varphi_{1}\rangle$ and $|\varphi_{-1}\rangle$
owing to the numerical breaking of particle--hole symmetry.
Since the hybridization of the Majorana zero modes is negligibly small,
$|\chi_{1}\rangle$ and $|\chi_{2}\rangle$ can be regarded as
superpositions of the two Majorana zero-mode eigenvectors.

We present a practical procedure for obtaining $|\varphi_{1}\rangle$ and
$|\varphi_{-1}\rangle$ from $|\chi_{1}\rangle$ and $|\chi_{2}\rangle$
not satisfying $|\chi_{2}\rangle = \Xi|\chi_{1}\rangle$.
In the first step, we define $|\chi_{1\pm}\rangle$ and
$|\chi_{2\pm}\rangle$ as
\begin{align}
  |\chi_{l\pm}\rangle =
  c_{l\pm} \left(|\chi_{l}\rangle \pm \Xi|\chi_{l}\rangle\right) ,
\end{align}
where $l = 1$, $2$ and $c_{l\pm}$ is a normalization constant.
They satisfy $\Xi|\chi_{l\pm}\rangle = \pm |\chi_{l\pm}\rangle$,
which indicates that they can be regarded as the zero energy eigenvectors
corresponding to the Majorana zero modes.
Since $|\chi_{1}\rangle$ and $|\chi_{2}\rangle$ are formed only by
the two Majorana zero-mode eigenvectors,
$|\chi_{1+}\rangle$ and $|\chi_{2+}\rangle$ must be equivalent.
We rewrite them as $|\varphi_{+}\rangle$.
Similarly, we rewrite $|\chi_{1-}\rangle$ and $|\chi_{2-}\rangle$
as $|\varphi_{-}\rangle$.
We can construct Majorana operators using $|\varphi_{+}\rangle$ and
$|\varphi_{-}\rangle$ as shown in Appendix~A.
Note that $|\varphi_{+}\rangle$ and $|\varphi_{-}\rangle$ are localized
near opposite ends.

In the second step, we superpose $|\varphi_{+}\rangle$
and $|\varphi_{-}\rangle$
to form $|\varphi_{1}\rangle$ and $|\varphi_{-1}\rangle$ as
\begin{align}
    \label{eq:def-varphi_1}
  |\varphi_{1}\rangle
 & = \frac{1}{\sqrt{2}}\left(|\varphi_{+}\rangle + |\varphi_{-}\rangle\right) ,
    \\
    \label{eq:def-tildevarphi_1}
  |\varphi_{-1}\rangle
 & = \frac{1}{\sqrt{2}}\left(|\varphi_{+}\rangle - |\varphi_{-}\rangle\right) ,
\end{align}
which satisfy $\Xi|\varphi_{1}\rangle = |\varphi_{-1}\rangle$.
Note that the roles of $|\varphi_{1}\rangle$ and
$|\varphi_{-1}\rangle$ are reversed
if we change the sign of $|\varphi_{-}\rangle$.
This means that we must fix the signs of $|\varphi_{+}\rangle$
and $|\varphi_{-}\rangle$ in an appropriate manner.
We do this by fixing the signs of $|\varphi_{+}\rangle$ and
$|\varphi_{-}\rangle$ so that the resulting $|\varphi_{\pm 1}\rangle$ obeys
relations that are satisfied by $|\varphi_{\pm 1}\rangle$
when $N$ is sufficiently small.
It is difficult to obtain $E_{1}$ precisely.
Fortunately, the precise value is not important
for solving a scattering problem in the presence of the charging effect
as long as $E_{1}/t$ is sufficiently small.
Thus, we set $E_{1}/t = 1.0 \times 10^{-14}$ in actual simulations.

In the remainder of this section, we briefly consider
the energy gap and the critical magnetic field $h_{\rm c}$
of the topological superconductor using the Fourier transform
of $h_{\rm TS}$ given by
\begin{align}
 & h_{\rm TS}(k) =
      \nonumber \\
 & \left[ \begin{array}{cccc}
             h-m(k) & -i\Lambda(k) & 0 & \Delta \\
             i\Lambda(k) & -h-m(k) & -\Delta & 0 \\
             0 & -\Delta & -h+m(k) & i\Lambda(k) \\
             \Delta & 0 & -i\Lambda(k) & h+m(k)
         \end{array}
  \right] ,
\end{align}
where $m(k) = \mu_{\rm S}+2t\cos ka$ and $\Lambda(k)=2\lambda \sin ka$.
The eigenvalues of energy $E$ satisfy ${\rm det}\{h_{\rm TS}(k)-E\}=0$,
which is reduced to
\begin{align}
 & E^{4}-2\Gamma_{+}(k)E^{2}+\Gamma_{-}(k)^{2}+4\Delta^{2}\Lambda(k)^{2} = 0 ,
\end{align}
where $\Gamma_{\pm}(k) = m(k)^{2}+\Delta^{2}\pm(h^{2}+\Lambda(k)^{2})$.
The spectrum of the system is determined as
\begin{align}
  E_{\pm}(k) = \sqrt{\Gamma_{+}(k)
               \pm2\sqrt{m(k)^{2}(h^{2}+\Lambda(k)^{2})+h^{2}\Delta^{2}}} .
\end{align}
The energy gap of the system is defined as
\begin{align}
  E_{\rm g} = \min_{k} \{E_{-}(k)\} .
\end{align}
At $h = h_{\rm c}$, where the energy gap vanishes,
$h_{\rm TS}$ has an eigenstate at zero energy.
The above algebraic equation has the solution $E = 0$ only if
$\Gamma_{-}(k)^{2} + 4\Delta^{2}\Lambda(k)^{2} = 0$.
When $\mu_{\rm S} < 0$, this holds at $k = 0$, resulting in
\begin{align}
   \label{eq:def-h_c}
  h_{\rm c} = \sqrt{(\mu_{\rm S}+2t)^{2}+\Delta^{2}} .
\end{align}
The system becomes topologically nontrivial when $h > h_{\rm c}$.
It is convenient to introduce an approximate expression for $E_{\rm g}$
given by $E_{\rm g}^{\rm A} = | h-h_{\rm c}|$.
The right-hand side of this equation is equal to $E_{-}(k)$ at $k = 0$.
In the case with $\mu_{\rm S} = -1.9 t$, $\lambda = 0.3 t$,
and $\Delta = 0.1 t$, $E_{\rm g}^{\rm A}$ is exactly equal to $E_{\rm g}$
in the range of $0.63 \lesssim h/h_{\rm c} \lesssim 1.62$.

\section{Boundary Potential}

The boundary potential~\cite{Affleck,takane2} describes the effect of
the superconductor on the normal metal lead at a given energy $E$.
We obtain its components using a functional integration technique
in the absence of the charging effect.
Using Eqs.~(\ref{eq:def-d}) and (\ref{eq:def-d-dag})
with Eqs.~(\ref{eq:elements-varphi})--(\ref{eq:V-dag}),
we can express $H_{\rm T}$ in terms of $\{d_{n}\}$ and $\{d_{n}^{\dagger}\}$.
In the Matsubara representation, the action for the topological superconductor
including the coupling with the normal metal lead is $S = S_{0}+S_{\rm T}$ with
\begin{align}
  S_{0}
 & = \sum_{\omega}\sum_{n=1}^{2N}
     \bar{d}_{n}(\omega)\left(-i\omega + E_{n}\right)d_{n}(\omega) ,
       \\
  S_{\rm T}
 & = \sum_{\omega}\sum_{n=1}^{2N}\sum_{\sigma=\uparrow,\downarrow}
       \nonumber \\
 & \hspace{-5mm}
     \times
     \Biggl\{
       \biggl[ -t_{1}\left( u_{n\sigma}(1)\bar{f}_{1\sigma}(\omega)
                           -v_{n\sigma}(1)f_{1\sigma}(-\omega) \right)
       \nonumber \\
 & \hspace{-4mm}
              -t_{2}\left( e^{i\phi}u_{n\sigma}(N)\bar{f}_{M\sigma}(\omega)
                          -e^{-i\phi}v_{n\sigma}(N)f_{M\sigma}(-\omega) \right)
       \biggr] d_{n}(\omega)
       \nonumber \\
 & \hspace{-1mm}
     + \bar{d}_{n}(\omega)
       \biggl[ -t_{1}\left( u_{n\sigma}(1)f_{1\sigma}(\omega)
                           -v_{n\sigma}(1)\bar{f}_{1\sigma}(-\omega)
                   \right)
       \nonumber \\
 & \hspace{-4mm}
             -t_{2}\left( e^{-i\phi}u_{n\sigma}(N)f_{M\sigma}(\omega)
                    -e^{i\phi}v_{n\sigma}(N)\bar{f}_{M\sigma}(-\omega)
                   \right)
       \biggr]
     \Biggr\} ,
\end{align}
where $\omega$ represents the fermion Matsubara frequency,
and $\{\bar{d}_{n}(\omega)\}$, $\{d_{n}(\omega)\}$,
$\{\bar{f}_{n}(\omega)\}$, and $\{f_{n}(\omega)\}$
are sets of Grassmann variables.
Hereafter, we assume that $\{u_{n\sigma}(j)\}$ and $\{v_{n\sigma}(j)\}$
are real functions of $j$.

We obtain the boundary potential from $S = S_{0}+S_{\rm T}$ by integrating out
the fermion degrees of freedom in the topological superconductor.
Since $E_{1} > 0$, the nonlocal state is empty in the ground state.
For the ground state with the occupied nonlocal state
in the presence of the charging effect,
we need to modify the derivation as described in the next section.
The effective action $S_{\rm B}$ corresponding to the boundary potential
is defined as
\begin{align}
    \label{eq:def-S_B}
  \exp\left( -S_{\rm B} \right)
  = \frac{\int \prod_{\omega, n}D\bar{d}_{n}(\omega)Dd_{n}(\omega)
          \exp\left(-S_{0}-S_{\rm T}\right)}
         {\int \prod_{\omega, n}D\bar{d}_{n}(\omega)Dd_{n}(\omega)
          \exp\left(-S_{0}\right)} .
\end{align}
After carrying out the integrations, we find
\begin{align}
        \label{eq:def-S-B-Mat}
   S_{\rm B}
 & = \frac{1}{2}\sum_{\omega}\sum_{\zeta, \zeta' = 1, M}
     \bar{F}_{\zeta}(\omega)V_{\zeta \zeta'}(\omega)F_{\zeta'}(\omega) ,
\end{align}
where
\begin{align}
   F_{1}(\omega)
 & = \left[ f_{1\uparrow}(\omega) \, f_{1\downarrow}(\omega) \,
            \bar{f}_{1\uparrow}(-\omega) \,
            \bar{f}_{1\downarrow}(-\omega)
     \right]^\mathsf{T} ,
     \\
   \bar{F}_{1}(\omega)
 & = \left[ \bar{f}_{1\uparrow}(\omega) \, \bar{f}_{1\downarrow}(\omega) \,
            f_{1\uparrow}(-\omega) \, f_{1\downarrow}(-\omega)
          \right] ,
     \\
   F_{M}(\omega)
 & = \left[ f_{M\uparrow}(\omega) \, f_{M\downarrow}(\omega) \,
            \bar{f}_{M\uparrow}(-\omega) \,
            \bar{f}_{M\downarrow}(-\omega)
     \right]^\mathsf{T} ,
     \\
   \bar{F}_{M}(\omega)
 & = \left[ \bar{f}_{M\uparrow}(\omega) \,
            \bar{f}_{M1\downarrow}(\omega) \,
            f_{M\uparrow}(-\omega) \, f_{M\downarrow}(-\omega)
          \right] .
\end{align}
The $4 \times 4$ matrices $V_{1 1}(\omega)$, $V_{M M}(\omega)$,
$V_{1 M}(\omega)$, and $V_{M 1}(\omega)$ constitute
the boundary potential in the Matsubara representation.
They are expressed as
\begin{align}
 & V_{\zeta \zeta'}(\omega)
         \nonumber \\
 & = \left[ \begin{array}{cccc}
              V_{\zeta \zeta'}^{e\uparrow, e\uparrow}(\omega)
              & V_{\zeta \zeta'}^{e\uparrow, e\downarrow}(\omega)
              & V_{\zeta \zeta'}^{e\uparrow, h\uparrow}(\omega)
              & V_{\zeta \zeta'}^{e\uparrow, h\downarrow}(\omega)
                  \\
              V_{\zeta \zeta'}^{e\downarrow, e\uparrow}(\omega)
              & V_{\zeta \zeta'}^{e\downarrow, e\downarrow}(\omega)
              & V_{\zeta \zeta'}^{e\downarrow, h\uparrow}(\omega)
              & V_{\zeta \zeta'}^{e\downarrow, h\downarrow}(\omega)
                  \\
              V_{\zeta \zeta'}^{h\uparrow, e\uparrow}(\omega)
              & V_{\zeta \zeta'}^{h\uparrow, e\downarrow}(\omega)
              & V_{\zeta \zeta'}^{h\uparrow, h\uparrow}(\omega)
              & V_{\zeta \zeta'}^{h\uparrow, h\downarrow}(\omega)
                  \\
              V_{\zeta \zeta'}^{h\downarrow, e\uparrow}(\omega)
              & V_{\zeta \zeta'}^{h\downarrow, e\downarrow}(\omega)
              & V_{\zeta \zeta'}^{h\downarrow, h\uparrow}(\omega)
              & V_{\zeta \zeta'}^{h\downarrow, h\downarrow}(\omega)
            \end{array}
     \right] .
\end{align}
The role of each matrix element of $V_{\zeta \zeta'}(\omega)$
is straightforward.
For example, $V_{1 M}^{e\uparrow, h\downarrow}(\omega)$
describes the scattering of a spin-down hole at the $M$th site
to a spin-up electron at the $1$st site.
We can express $V_{\zeta \zeta'}^{\eta\sigma, \eta'\sigma'}(\omega)$
with $\eta, \eta' = e, h$ and $\sigma, \sigma' = \uparrow, \downarrow$
in terms of $\{E_{n}\}$ and $\{u_{n\sigma}(1)\}$, $\{v_{n\sigma}(1)\}$,
$\{u_{n\sigma}(N)\}$, and $\{v_{n\sigma}(N)\}$, as shown in Appendix~B.

Let us consider a scattering problem in the interferometer
when an electron with energy $E$ is incident from the left lead,
ignoring the charging effect.
The effect of the topological superconductor on the incident electron
is described by the boundary potential consisting of the $4 \times 4$ matrices
$V_{\zeta \zeta'}(E)$ with $\zeta, \zeta' = 1, M$ defined by
\begin{align}
  V_{\zeta \zeta'}(E)
  = \left. V_{\zeta \zeta'}(\omega)\right|_{i\omega \to E+ i\delta} ,
\end{align}
where $\delta$ is a positive infinitesimal.
For example, $V_{1 M}^{e\sigma, e\sigma'}(E)$ is expressed as
\begin{align}
     \label{eq:ex-V12}
  V_{1 M}^{e\sigma, e\sigma'}(E)
 & = -e^{-i\phi}t_{1}t_{2}\sum_{n=1}^{2N}
      \nonumber \\
 & \hspace{-12mm}
      \times
      \left( \frac{u_{n\sigma}(1)u_{n\sigma'}(N)}{-E -i\delta +E_{n}}
            - \frac{v_{n\sigma}(1)v_{n\sigma'}(N)}{E + i\delta + E_{n}}
      \right) .
\end{align}
Once the eigenvalues and eigenvectors of $h_{\rm TS}$ are obtained numerically,
we can straightforwardly calculate the elements of the matrices constituting
the boundary potential for any $E$ and $\phi$.

To express the effective BdG Hamiltonian
for this scattering problem at energy $E$,
we introduce the four-component state vector
for the $j$th site in the normal metal lead as
\begin{align}
  |j \rangle
  = \Bigl[ |j \rangle_{\uparrow}^{e} \hspace{1.0mm}
             |j \rangle_{\downarrow}^{e} \hspace{1.0mm}
             |j \rangle_{\uparrow}^{h} \hspace{1.0mm}
             |j \rangle_{\downarrow}^{h}
    \Bigr] .
\end{align}
With $|j \rangle$ and $\langle j| \equiv |j \rangle^{\dagger}$,
the effective BdG Hamiltonian is given by
\begin{align}
        \label{eq:H_eff}
  H_{\rm eff}
 & = \sum_{j = - \infty}^{\infty}
     \bigl( |j+1 \rangle h_{t'} \langle j| + |j \rangle h_{t'} \langle j+1|
          + |j \rangle h_{\mu} \langle j| \bigr)
       \nonumber \\
 & \hspace{7mm}
     + |1 \rangle V_{1 1}(E) \langle 1| + |M \rangle V_{M M}(E) \langle M|
       \nonumber \\
 & \hspace{7mm}
     + |1 \rangle V_{1 M}(E) \langle M| + |M \rangle V_{M 1}(E) \langle 1| ,
\end{align}
where
\begin{align}
  h_{t'} & = \left[ \begin{array}{cccc}
                      -t' & 0 & 0 & 0 \\
                      0 & -t' & 0 & 0 \\
                      0 & 0 & t' & 0 \\
                      0 & 0 & 0 & t'
                    \end{array}
             \right] ,
     \\
  h_{\mu} & = \left[ \begin{array}{cccc}
                       -\mu_{\rm N} & 0 & 0 & 0 \\
                       0 & -\mu_{\rm N} & 0 & 0 \\
                       0 & 0 & \mu_{\rm N} & 0 \\
                       0 & 0 & 0 & \mu_{\rm N}
                     \end{array}
              \right] .
\end{align}
A scattering state is written as
\begin{align}
  |\Psi \rangle = \sum_{j = -\infty}^{\infty}
                  |j \rangle \cdot \mib{\psi}(j)
\end{align}
with
\begin{align}
  \mib{\psi}(j)
  = \left[ \psi_{\uparrow}^{e}(j) \, \psi_{\downarrow}^{e}(j) \,
           \psi_{\uparrow}^{h}(j) \, \psi_{\downarrow}^{h}(j)
    \right]^\mathsf{T} .
\end{align}

Let us introduce the reflection coefficients $r_{\sigma \sigma'}^{e}$ and
$r_{\sigma \sigma'}^{h}$, and the transmission coefficients
$t_{\sigma \sigma'}^{e}$ and $t_{\sigma \sigma'}^{h}$.
Here, $r_{\sigma \sigma'}^{e}$ ($r_{\sigma \sigma'}^{h}$) represents
the reflection coefficient describing the process
in which an incident electron with spin-$\sigma'$ is reflected back to
the left lead as an electron (a hole) with spin-$\sigma$,
and $t_{\sigma \sigma'}^{e}$ ($t_{\sigma \sigma'}^{h}$) represents
the transmission coefficient describing the process
in which an incident electron with spin-$\sigma'$ is transmitted to
the right lead as an electron (a hole) with spin-$\sigma$.
Scattering states in the left and right leads are written in terms of
these reflection and transmission coefficients.
When a spin-up electron with energy $E$ is incident from
the left lead, the scattering state is written as
\begin{align}
 \mib{\psi}(j)
 = \left[ \begin{array}{c}
            e^{ik(x_{j}-x_{1})}
              + r_{\uparrow\uparrow}^{e} e^{-ik(x_{j}-x_{1})} \\
            r_{\downarrow\uparrow}^{e} e^{-ik(x_{j}-x_{1})} \\
            \sqrt{\frac{\sin ka}{\sin qa}} \, r_{\uparrow\uparrow}^{h}
              e^{iq(x_{j}-x_{1})} \\
            \sqrt{\frac{\sin ka}{\sin qa}} \, r_{\downarrow\uparrow}^{h}
              e^{iq(x_{j}-x_{1})}
          \end{array}
   \right]
\end{align}
for $j \le 1$ and 
\begin{align}
 \mib{\psi}(j)
 = \left[ \begin{array}{c}
            t_{\uparrow\uparrow}^{e} e^{ik(x_{j}-x_{M})} \\
            t_{\downarrow\uparrow}^{e} e^{ik(x_{j}-x_{M})} \\
            \sqrt{\frac{\sin ka}{\sin qa}} \, t_{\uparrow\uparrow}^{h}
              e^{-iq(x_{j}-x_{M})} \\
            \sqrt{\frac{\sin ka}{\sin qa}} \, t_{\downarrow\uparrow}^{h}
              e^{-iq(x_{j}-x_{M})}
          \end{array}
   \right]
\end{align}
for $j \ge M$, where $x_{j} = ja$,
and the wave number $k$ in the electron waves is determined by
the condition $E = -2t'\cos(ka)-\mu_{\rm N} \equiv E^{e}(k)$
with $\sin(ka) > 0$.
The hole waves appear when we can determine the wave number $q$
such that $E = 2t'\cos(qa)+\mu_{\rm N} \equiv E^{h}(q)$
with $\sin(qa) > 0$.
Here, $\sin(ka) > 0$ results from the condition
that the velocity of an incoming electron must be positive in the left lead:
$v_{\rm in}^{e}  = \partial E^{e}/\partial k > 0$.
Similarly, $\sin(qa) > 0$ results from the condition
that the velocity of an outgoing hole must be negative in the left lead:
$v_{\rm out}^{h} = \partial E^{h}/\partial q < 0$.

The transmission and reflection coefficients
are obtained as follows.~\cite{Ando}
When a spin-up electron with energy $E$ is incident from the left lead,
the equations of $\psi_{\sigma}^{e}(j)$ and $\psi_{\sigma}^{h}(j)$
at $j = 1$ and $M$ are described as
\begin{align}
      \label{eq:set1}
 & \Bigl[E+\mu_{\rm N}+t'e^{ika}-V_{11}^{e\sigma,e\sigma}(E)\Bigr]
   \psi_{\sigma}^{e}(1)
       \nonumber \\
 & +t'\psi_{\sigma}^{e}(2)-V_{11}^{e\sigma,e\bar{\sigma}}(E)
                           \psi_{\bar{\sigma}}^{e}(1)
       \nonumber \\
 & -\sum_{\sigma'}
    \Bigl[ V_{1M}^{e\sigma,e\sigma'}(E)\psi_{\sigma'}^{e}(M)
         + V_{11}^{e\sigma,h\sigma'}(E)\psi_{\sigma'}^{h}(1)
       \nonumber \\
 & \hspace{8mm}
         + V_{1M}^{e\sigma,h\sigma'}(E)\psi_{\sigma'}^{h}(M)
    \Bigr]
    = 2it'\sin(ka) \times \delta_{\sigma, \uparrow} ,
       \\
      \label{eq:set2}
 & \Bigl[E-\mu_{\rm N}-t'e^{-iqa}-V_{11}^{h\sigma,h\sigma}(E)\Bigr]
   \psi_{\sigma}^{h}(1)
       \nonumber \\
 & -t'\psi_{\sigma}^{h}(2)-V_{11}^{h\sigma,h\bar{\sigma}}(E)
                           \psi_{\bar{\sigma}}^{h}(1)
       \nonumber \\
 & -\sum_{\sigma'}
    \Bigl[ V_{11}^{h\sigma,e\sigma'}(E)\psi_{\sigma'}^{e}(1)
         + V_{1M}^{h\sigma,e\sigma'}(E)\psi_{\sigma'}^{e}(M)
       \nonumber \\
 & \hspace{8mm}
         + V_{1M}^{h\sigma,h\sigma'}(E)\psi_{\sigma'}^{h}(M)
    \Bigr]
    = 0 ,
       \\
      \label{eq:set3}
 & \Bigl[E+\mu_{\rm N}+t'e^{ika}-V_{MM}^{e\sigma,e\sigma}(E)\Bigr]
   \psi_{\sigma}^{e}(M)
       \nonumber \\
 & +t'\psi_{\sigma}^{e}(M-1)-V_{MM}^{e\sigma,e\bar{\sigma}}(E)
                           \psi_{\bar{\sigma}}^{e}(M)
       \nonumber \\
 & -\sum_{\sigma'}
    \Bigl[ V_{M1}^{e\sigma,e\sigma'}(E)\psi_{\sigma'}^{e}(1)
         + V_{M1}^{e\sigma,h\sigma'}(E)\psi_{\sigma'}^{h}(1)
       \nonumber \\
 & \hspace{8mm}
         + V_{MM}^{e\sigma,h\sigma'}(E)\psi_{\sigma'}^{h}(M)
    \Bigr]
    = 0 ,
       \\
      \label{eq:set4}
 & \Bigl[E-\mu_{\rm N}-t'e^{-iqa}-V_{MM}^{h\sigma,h\sigma}(E)\Bigr]
   \psi_{\sigma}^{h}(M)
       \nonumber \\
 & -t'\psi_{\sigma}^{h}(M-1)-V_{MM}^{h\sigma,h\bar{\sigma}}(E)
                           \psi_{\bar{\sigma}}^{h}(M)
       \nonumber \\
 & -\sum_{\sigma'}
    \Bigl[ V_{M1}^{h\sigma,e\sigma'}(E)\psi_{\sigma'}^{e}(1)
         + V_{MM}^{h\sigma,e\sigma'}(E)\psi_{\sigma'}^{e}(M)
       \nonumber \\
 & \hspace{8mm}
         + V_{M1}^{h\sigma,h\sigma'}(E)\psi_{\sigma'}^{h}(1)
    \Bigr]
    = 0 ,
\end{align}
where $\bar{\uparrow} = \downarrow$ and $\bar{\downarrow} = \uparrow$.
At $2 \le j \le M-1$,
$\psi_{\sigma}^{e}(j)$ and $\psi_{\sigma}^{h}(j)$ satisfy
\begin{align}
      \label{eq:set5}
  \left(E+\mu_{\rm N}\right)\psi_{\sigma}^{e}(j)
   +t'\psi_{\sigma}^{e}(j-1)+t'\psi_{\sigma}^{e}(j+1) = 0 ,
       \\
      \label{eq:set6}
  \left(E-\mu_{\rm N}\right)\psi_{\sigma}^{h}(j)
   -t'\psi_{\sigma}^{h}(j-1)-t'\psi_{\sigma}^{h}(j+1) = 0 ,
\end{align}
respectively.
Solving the set of Eqs.~(\ref{eq:set1})--(\ref{eq:set6}),
we can numerically determine $\psi_{\sigma}^{e}(1)$,
$\psi_{\sigma}^{h}(1)$, $\psi_{\sigma}^{e}(M)$, and $\psi_{\sigma}^{h}(M)$,
from which we can readily obtain the transmission and reflection coefficients.
For example, $r_{\uparrow \uparrow}^{e}$ and $t_{\downarrow \uparrow}^{h}$
are obtained as
\begin{align}
  r_{\uparrow \uparrow}^{e} & = \psi_{\uparrow}^{e}(1) -1 ,
     \\
  t_{\downarrow \uparrow}^{h} & = \sqrt{\frac{\sin qa}{\sin ka}}
                                  \psi_{\downarrow}^{h}(M) .
\end{align}

\section{Charging Effect on the Scattering Problem}

We consider the scattering problem under a given constraint
such as ${\rm G}_{2l} \leftrightarrow {\rm E}_{2l \pm 1}$
and ${\rm G}_{2l\pm 1} \leftrightarrow {\rm E}_{2l}$.
Andreev reflection processes are forbidden
because these processes increase or decrease $\mathcal{N}$ by two.
Therefore, an electron state is not mixed with a hole state.
As a result, the scattering processes of an incident electron
from the left lead can be described only in the electron space.
That is, we are allowed to consider only
the equations for $\psi_{\sigma}^{e}(j)$.
In the equations for $\psi_{\sigma}^{e}(j)$ at $j = 1$ and $M$,
we must exclude the terms with $V_{\zeta \zeta'}^{h\sigma,e\sigma'}(E)$,
which describe Andreev reflection processes.
Thus, we include only the terms with $V_{\zeta \zeta'}^{e\sigma,e\sigma'}(E)$,
which must be modified in the presence of the charging effect.
We express the modified components of the boundary potential as
$[V_{\zeta \zeta'}^{e\sigma,e\sigma'}(E)]_{\rm C}$.
Once these components of the boundary potential are given,
the equations of $\psi_{\sigma}^{e}(j)$ at $j = 1$ and $M$ are expressed as
\begin{align}
      \label{eq:set1+}
 & \Bigl[E+\mu_{\rm N}+t'e^{ika}
         -\bigl[V_{11}^{e\sigma,e\sigma}(E)\bigr]_{\rm C}
   \Bigr]\psi_{\sigma}^{e}(1)
   +t'\psi_{\sigma}^{e}(2)
       \nonumber \\
 & -\bigl[V_{11}^{e\sigma,e\bar{\sigma}}(E)\bigr]_{\rm C}
                           \psi_{\bar{\sigma}}^{e}(1)
   -\sum_{\sigma'}
    \bigl[V_{1M}^{e\sigma,e\sigma'}(E)\bigr]_{\rm C}\psi_{\sigma'}^{e}(M)
       \nonumber \\
 & \hspace{2mm}
    = 2it'\sin(ka) \times \delta_{\sigma, \uparrow} ,
       \\
      \label{eq:set3+}
 & \Bigl[E+\mu_{\rm N}+t'e^{ika}
         -\bigl[V_{MM}^{e\sigma,e\sigma}(E)\bigr]_{\rm C}
   \Bigr]\psi_{\sigma}^{e}(M)
   +t'\psi_{\sigma}^{e}(M-1)
       \nonumber \\
 & -\big[V_{MM}^{e\sigma,e\bar{\sigma}}(E)\bigr]_{\rm C}
    \psi_{\bar{\sigma}}^{e}(M)
   -\sum_{\sigma'}
    \bigl[V_{M1}^{e\sigma,e\sigma'}(E)\bigr]_{\rm C}\psi_{\sigma'}^{e}(1)
    = 0 ,
\end{align}
respectively.
The equations of $\psi_{\sigma}^{e}(j)$ at $2 \le j \le M-1$
are unchanged.
Solving the set of Eqs.~(\ref{eq:set5}), (\ref{eq:set1+}),
and (\ref{eq:set3+}), we can numerically determine
$\psi_{\sigma}^{e}(1)$ and $\psi_{\sigma}^{e}(M)$,
from which we can readily obtain the transmission and reflection coefficients.
%%%%%%%%%%%%%%%%%%
\begin{figure}[tbp]
\begin{tabular}{cc}
\begin{minipage}{0.5\hsize}
\begin{center}
\hspace{-9mm}
\includegraphics[height=2.4cm]{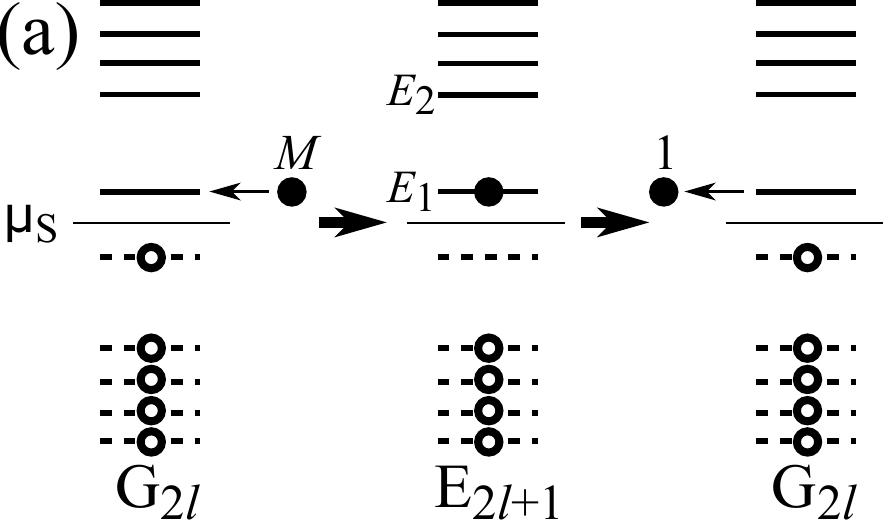}
\end{center}
\end{minipage}
\begin{minipage}{0.5\hsize}
\begin{center}
\hspace{-6mm}
\includegraphics[height=2.4cm]{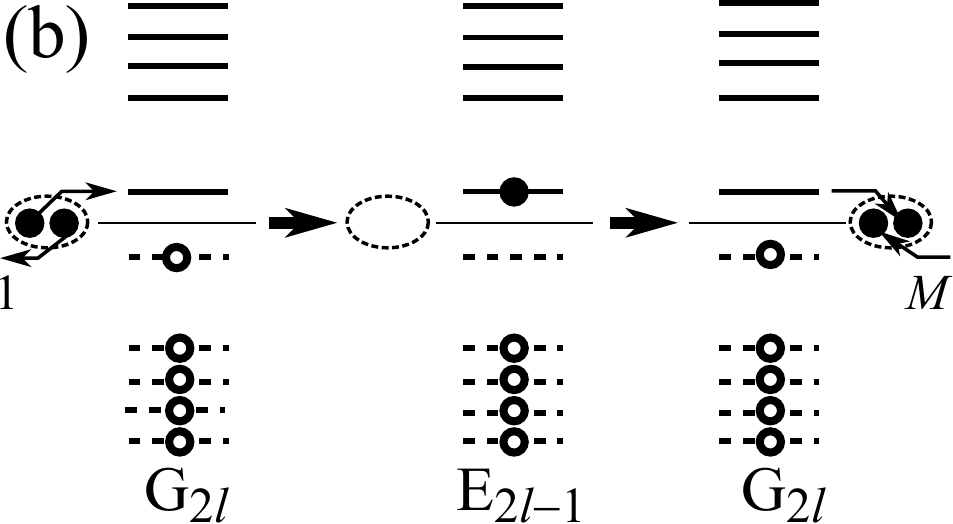}
\end{center}
\end{minipage}
\end{tabular}
\caption{
Schematics of scattering processes that transfer an electron at the $M$th site
to an electron at the 1st site when the ground state is ${\rm G}_{2l}$.
Only the processes involving the nonlocal state with $E_{1}$ are shown.
Process (a) is included in
$\bigl[V_{1 M}^{e\sigma, e\sigma'}(E)\bigr]_{+}^{\mathcal E}$,
whereas process (b) accompanying the splitting and recombination of
a Cooper pair is included in
$\bigl[V_{1 M}^{e\sigma, e\sigma'}(E)\bigr]_{-}^{\mathcal E}$.
}
\end{figure}
%%%%%%%%%%%%%%%%%%

We derive the modified components of the boundary potential
from those given in Sect.~3 in the absence of the charging effect.
Let us consider the case with the ground state ${\rm G}_{2l}$.
First, we decompose $V_{\zeta \zeta'}^{e\sigma,e\sigma'}(E)$ as
\begin{align}
 V_{\zeta \zeta'}^{e\sigma, e\sigma'}(E)
 = \bigl[V_{\zeta \zeta'}^{e\sigma, e\sigma'}(E)\bigr]_{+}
   + \bigl[V_{\zeta \zeta'}^{e\sigma, e\sigma'}(E)\bigr]_{-} ,
\end{align}
where the first (second) part consists of the terms
with a pole at $E > 0$ ($E < 0$).
For example, the two parts of $V_{1 M}^{e\sigma, e\sigma'}(E)$ are expressed as
\begin{align}
  \bigl[V_{1 M}^{e\sigma, e\sigma'}(E)\bigr]_{+}
 & = -e^{-i\phi}t_{1}t_{2}\sum_{n=1}^{2N}
      \frac{u_{n\sigma}(1)u_{n\sigma'}(N)}{-E -i\delta +E_{n}} ,
    \\
  \bigl[V_{1 M}^{e\sigma, e\sigma'}(E)\bigr]_{-}
 & = e^{-i\phi}t_{1}t_{2}\sum_{n=1}^{2N}
      \frac{v_{n\sigma}(1)v_{n\sigma'}(N)}{E + i\delta + E_{n}} .
\end{align}
Each term in $\bigl[V_{1 M}^{e\sigma, e\sigma'}(E)\bigr]_{+}$ describes
the scattering process,
in which an electron at the $M$th site tunnels into the superconductor
and subsequently tunnels to the 1st site [see Fig.~2(a)].
Each term in $\bigl[V_{1 M}^{e\sigma, e\sigma'}(E)\bigr]_{-}$ describes
the scattering process,
in which an electron in the superconductor tunnels to the 1st site, and
subsequently, an electron at the $M$th site tunnels into the superconductor,
accompanied by the splitting and recombination
of a Cooper pair [see Fig.~2(b)].
That is, $\bigl[V_{1 M}^{e\sigma, e\sigma'}(E)\bigr]_{+}$
($\bigl[V_{1 M}^{e\sigma, e\sigma'}(E)\bigr]_{-}$) describes
the scattering processes in which $\mathcal{N}$ increases (decreases) by one
in their intermediate states.
Second, we take account of the charging energy
$\delta U \equiv U(2l \pm 1) - U(2l)$
under the constraint ${\rm G}_{2l} \leftrightarrow {\rm E}_{2l \pm 1}$,
which is common to all bogolon states.
Adding this to $E_{n}$ as $E_{n}+\delta U$, we find
\begin{align}
    \label{eq:V_1M_+E}
  \bigl[V_{1 M}^{e\sigma, e\sigma'}(E)\bigr]_{+}^{\mathcal E}
 & = -e^{-i\phi}t_{1}t_{2}\sum_{n=1}^{2N}
      \frac{u_{n\sigma}(1)u_{n\sigma'}(N)}{-E -i\delta +E_{n}+\delta U} ,
    \\
    \label{eq:V_1M_-E}
  \bigl[V_{1 M}^{e\sigma, e\sigma'}(E)\bigr]_{-}^{\mathcal E}
 & = e^{-i\phi}t_{1}t_{2}\sum_{n=1}^{2N}
      \frac{v_{n\sigma}(1)v_{n\sigma'}(N)}{E + i\delta + E_{n}+\delta U} .
\end{align}
Here, the superscript $\mathcal E$ represents that
$\mathcal{N}$ in the ground state is even.
We conclude that
\begin{align}
  [V_{\zeta \zeta'}^{e\sigma,e\sigma'}(E)]_{\rm C}
  = \bigl[V_{\zeta \zeta'}^{e\sigma, e\sigma'}(E)\bigr]_{+}^{\mathcal E}
\end{align}
under the constraint ${\rm G}_{2l} \leftrightarrow {\rm E}_{2l + 1}$
and
\begin{align}
  [V_{\zeta \zeta'}^{e\sigma,e\sigma'}(E)]_{\rm C}
  = \bigl[V_{\zeta \zeta'}^{e\sigma, e\sigma'}(E)\bigr]_{-}^{\mathcal E}
\end{align}
under the constraint ${\rm G}_{2l} \leftrightarrow {\rm E}_{2l - 1}$.

Let us next consider the case with the ground state ${\rm G}_{2l \pm 1}$.
To describe ${\rm G}_{2l \pm 1}$ with $\langle d_{1}^\dagger d_{1}\rangle = 1$,
we must modify the boundary potential given in Sect.~3.
The charging effect causes $\langle d_{1}^\dagger d_{1}\rangle = 1$
in the ground state when the energy of the occupied nonlocal state,
$E_{1}+U(2l \pm 1)$, is smaller than that in the vacant case, $U(2l)$.
The corresponding excitation energy is $\delta U - E_{1}$,
where $\delta U \equiv U(2l) - U(2l\pm 1)$.
We thus replace the term $E_{1}d_{1}^{\dagger}d_{1}$ in $H_{\rm TS}$
with $(\delta U - E_{1})\tilde{d}_{1}^{\dagger}\tilde{d}_{1}$,
where $\tilde{d}_{1}^{\dagger} = d_{1}$ and $\tilde{d}_{1} = d_{1}^{\dagger}$.
The charging effect on the other bogolon states is considered later.
Carrying out the procedure described in Sect.~3,
we obtain the boundary potential in ${\rm G}_{2l\pm 1}$.
The resulting boundary potential is equivalent to that obtained
by replacing $E_{1}$ with $-E_{1}+\delta U$
and exchanging $u_{1\sigma}(j)$ and $v_{1\sigma}(j)$ for $j = 1, N$
in the boundary potential given in Sect.~3.
For example, $V_{1 M}^{e\sigma, e\sigma'}(E)$ is expressed as
\begin{align}
     \label{eq:ex-V12-CE2}
  V_{1 M}^{e\sigma, e\sigma'}(E)
 & = -e^{-i\phi}t_{1}t_{2}
      \nonumber \\
 & \hspace{-16mm}
     \times
     \Bigg[\frac{v_{1\sigma}(1)v_{1\sigma'}(N)}{-E-i\delta-E_{1}+\delta U}
            - \frac{u_{1\sigma}(1)u_{1\sigma'}(N)}{E+i\delta-E_{1}+\delta U}
      \nonumber \\
 & \hspace{-12mm}
      +\sum_{n=2}^{2N}
      \left( \frac{u_{n\sigma}(1)u_{n\sigma'}(N)}{-E-i\delta+E_{n}}
            - \frac{v_{n\sigma}(1)v_{n\sigma'}(N)}{E+i\delta+E_{n}}
      \right) \Bigg] .
\end{align}
We separate $V_{\zeta \zeta'}^{e\sigma, e\sigma'}(E)$ into
$\bigl[V_{\zeta \zeta'}^{e\sigma, e\sigma'}(E)\bigr]_{+}$
and $\bigl[V_{\zeta \zeta'}^{e\sigma, e\sigma'}(E)\bigr]_{-}$,
where $\bigl[V_{\zeta \zeta'}^{e\sigma, e\sigma'}(E)\bigr]_{+}$
($\bigl[V_{\zeta \zeta'}^{e\sigma, e\sigma'}(E)\bigr]_{-}$) describes
the scattering processes in which $\mathcal{N}$ increases (decreases) by one
in their intermediate states.
We then replace $E_{n}$ with $E_{n}+\delta U$ for $n \ge 2$ to take account
of the charging effect on the corresponding bogolon states.
In the case of $V_{1 M}^{e\sigma, e\sigma'}(E)$, this results in
\begin{align}
    \label{eq:V_1M_+O}
  \bigl[V_{1 M}^{e\sigma, e\sigma'}(E)\bigr]_{+}^{\mathcal O}
 & = -e^{-i\phi}t_{1}t_{2}
     \Bigg[ \frac{v_{1\sigma}(1)v_{1\sigma'}(N)}{-E-i\delta-E_{1}+\delta U}
      \nonumber \\
 & \hspace{5mm}
      +\sum_{n=2}^{2N}
      \frac{u_{n\sigma}(1)u_{n\sigma'}(N)}{-E-i\delta+E_{n}+\delta U}
      \Bigg] ,
    \\
    \label{eq:V_1M_-O}
  \bigl[V_{1 M}^{e\sigma, e\sigma'}(E)\bigr]_{-}^{\mathcal O}
 & = e^{-i\phi}t_{1}t_{2}
     \Bigg[ \frac{u_{1\sigma}(1)u_{1\sigma'}(N)}{E+i\delta-E_{1}+\delta U}
      \nonumber \\
 & \hspace{5mm}
      +\sum_{n=2}^{2N}
      \frac{v_{n\sigma}(1)v_{n\sigma'}(N)}{E+i\delta+E_{n}+\delta U}
      \Bigg] ,
\end{align}
where the superscript $\mathcal O$ represents that
$\mathcal{N}$ in the ground state is odd.
We conclude that
\begin{align}
  [V_{\zeta \zeta'}^{e\sigma,e\sigma'}(E)]_{\rm C}
  = \bigl[V_{\zeta \zeta'}^{e\sigma, e\sigma'}(E)\bigr]_{+}^{\mathcal O}
\end{align}
under the constraint ${\rm G}_{2l - 1} \leftrightarrow {\rm E}_{2l}$
and
\begin{align}
  [V_{\zeta \zeta'}^{e\sigma,e\sigma'}(E)]_{\rm C}
  = \bigl[V_{\zeta \zeta'}^{e\sigma, e\sigma'}(E)\bigr]_{-}^{\mathcal O}
\end{align}
under the constraint ${\rm G}_{2l+1} \leftrightarrow {\rm E}_{2l}$.

\section{Numerical Results}

%%%%%%%%%%%%%%%%%%
\begin{figure}[btp]
\begin{center}
\includegraphics[height=5.0cm]{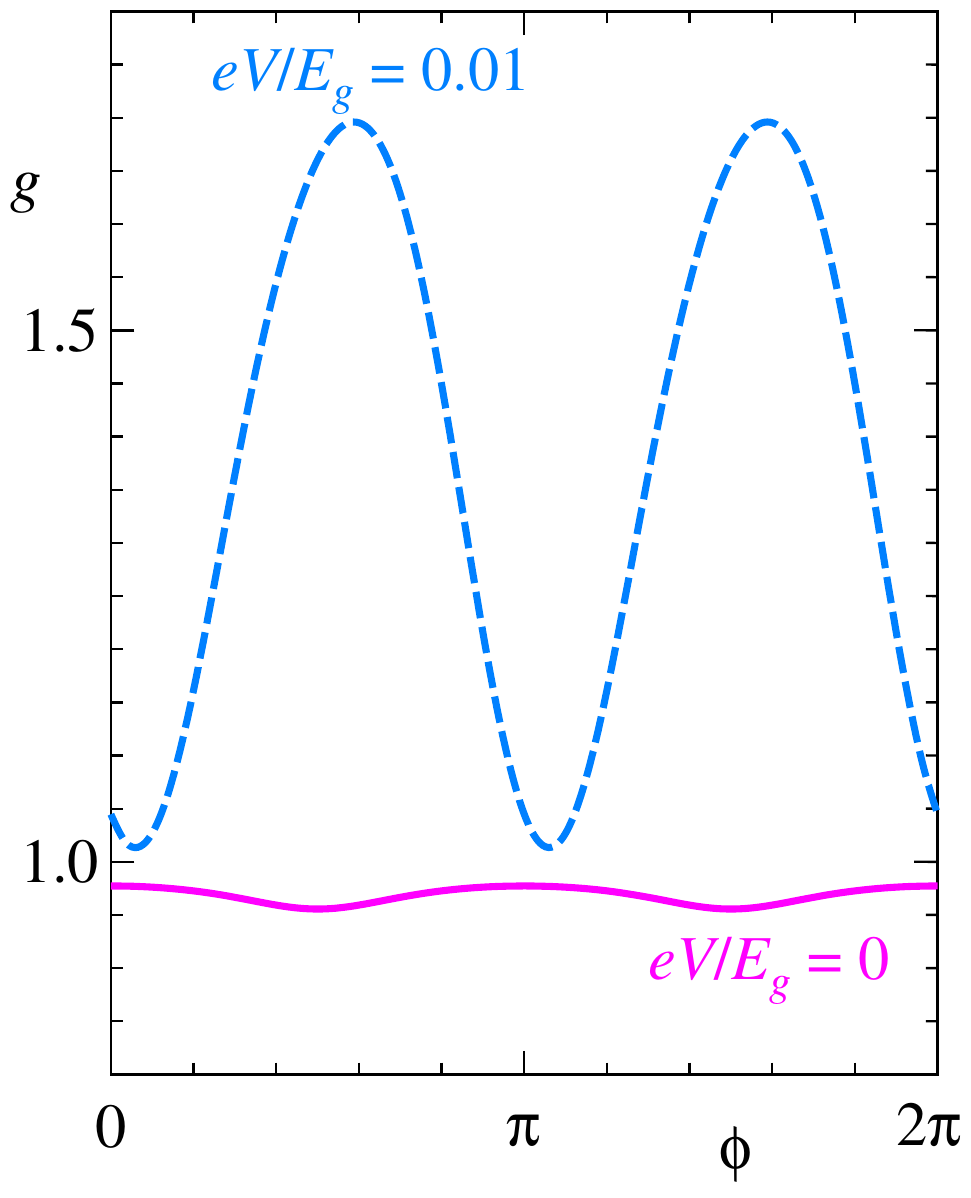}
\end{center}
\caption{
(Color online) Dimensionless nonlocal conductance $g(V)$ of the interferometer
in the absence of the charging effect at $eV/E_{g} = 0$ and $0.01$.
}
\end{figure}
%%%%%%%%%%%%%%%%%%
In this section, we numerically calculate the dimensionless nonlocal
conductance $g$ of the interferometer, which is defined
at zero temperature as~\cite{Lambert,Takane3,Anantram,Lesovik,Maiani}
\begin{align}
     \label{eq:def-nonl-g}
  g(V) = \sum_{\sigma, \sigma' = \uparrow, \downarrow}
         \left.\left( |t_{\sigma \sigma'}^{e}(E)|^{2}
               -|t_{\sigma \sigma'}^{h}(E)|^{2} \right)\right|_{E = eV} ,
\end{align}
as a function of $\phi$ in the absence and presence of the charging effect.
We focus on the low bias regime where $eV \ll E_{g}$ because we are interested
in the electron transport caused by electron teleportation.
The transmission probabilities in Eq.~(\ref{eq:def-nonl-g}) are determined
by solving the corresponding scattering problem.
In the presence of the charging effect, Andreev reflection processes are
forbidden and the second term of $g$ vanishes.
We set $N = 400$, $M = 440$, $t' = t$, $\Delta = 0.1 t$, $\lambda = 0.3 t$,
$t_{1} = t_{2} = 0.1 t$, $\mu_{\rm N} = 0$, and $\mu_{\rm S} = -1.9 t$.
Equation~(\ref{eq:def-h_c}) gives $h_{\rm c} = \sqrt{2} \Delta$.
The Zeeman field is fixed at $h = 1.35 h_{\rm c}$ so that the superconductor
becomes topologically nontrivial.
The energy gap of the superconductor is $E_{g} \approx 0.0495 t$.

%%%%%%%%%%%%%%%%%%
\begin{figure}[tbp]
\begin{tabular}{cc}
\begin{minipage}{0.5\hsize}
\begin{center}
\hspace{-5mm}
\includegraphics[height=5.5cm]{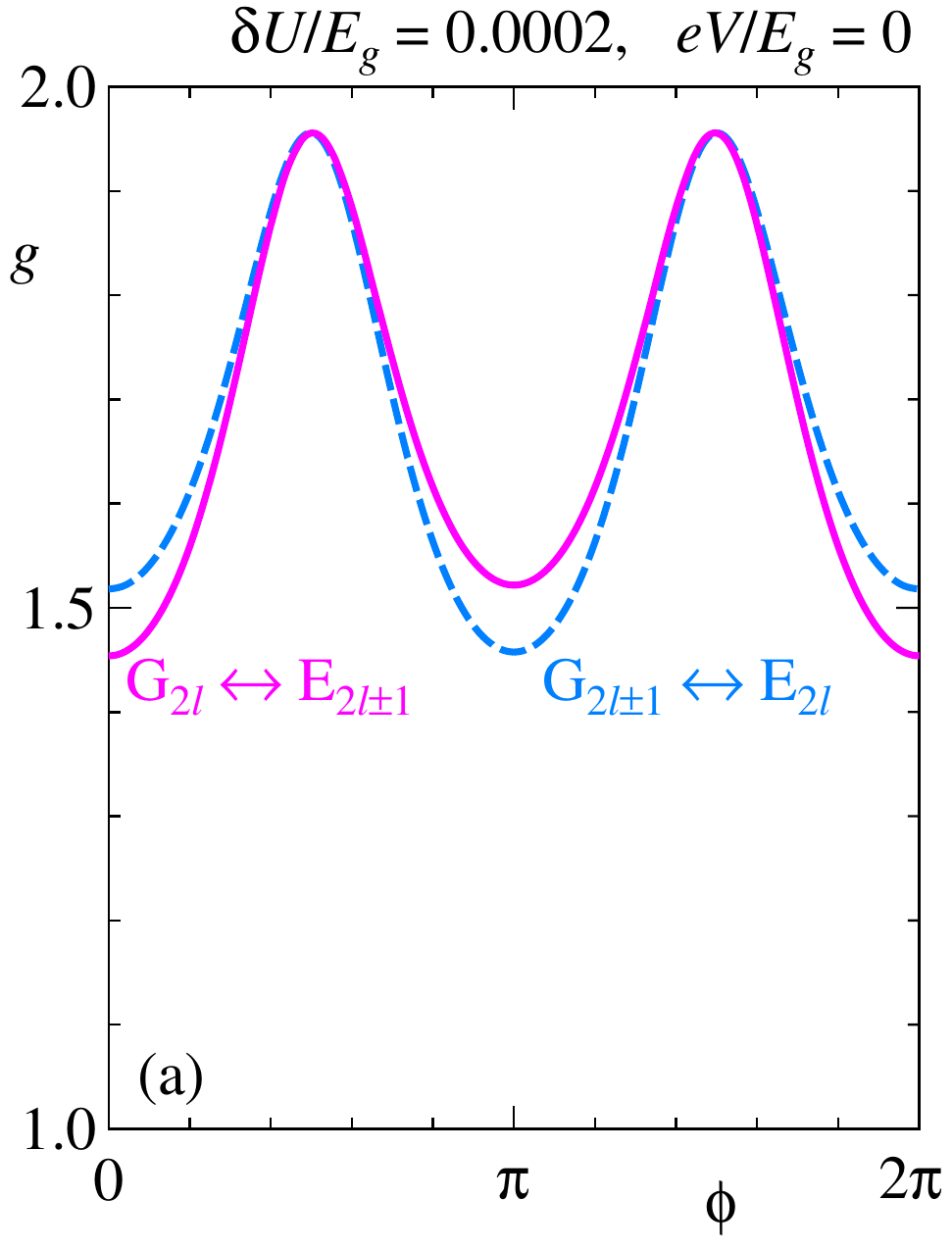}
\end{center}
\end{minipage}
\begin{minipage}{0.5\hsize}
\begin{center}
\hspace{-5mm}
\includegraphics[height=5.5cm]{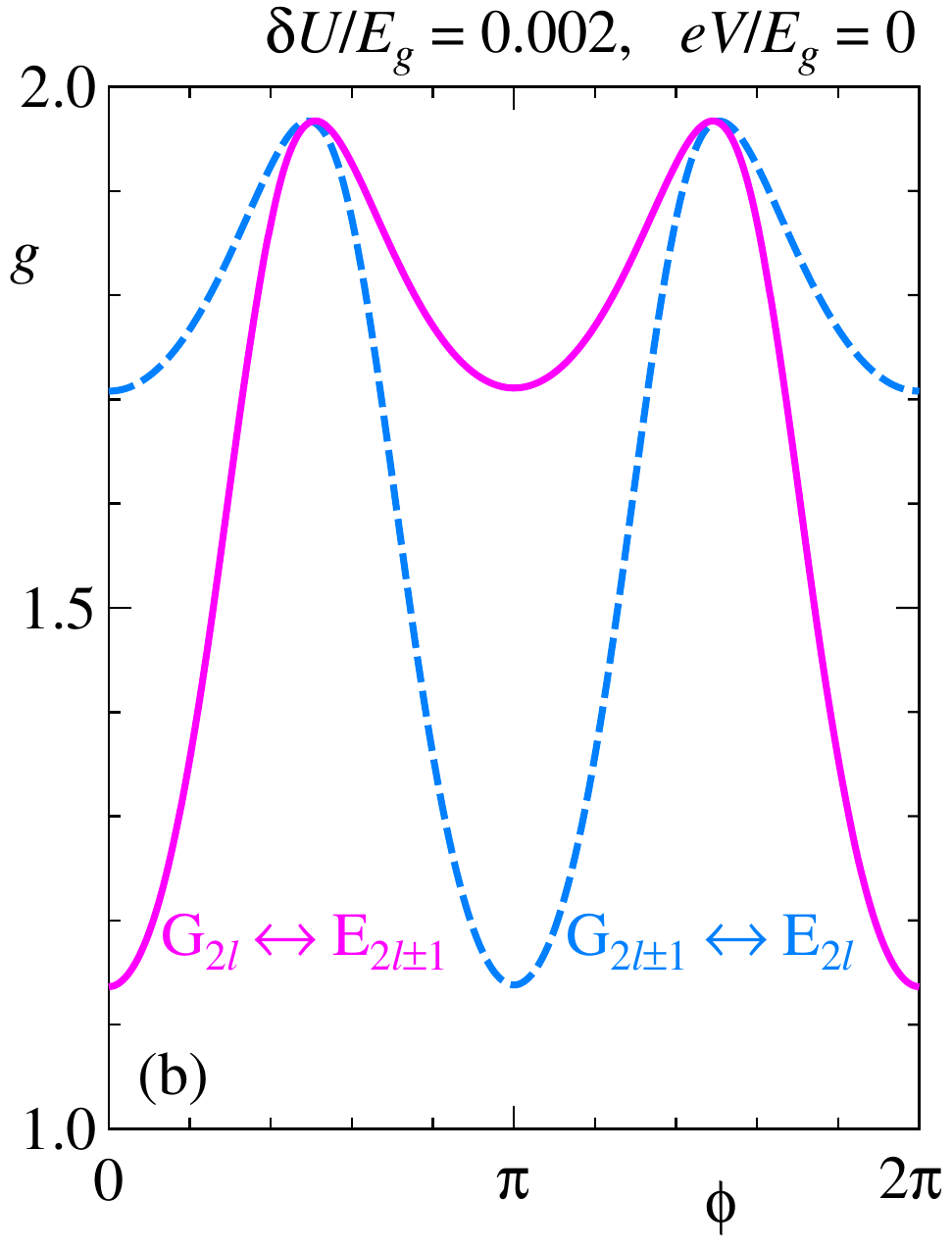}
\end{center}
\end{minipage}
\end{tabular}
\begin{tabular}{cc}
\begin{minipage}{0.5\hsize}
\begin{center}
\hspace{-5mm}
\includegraphics[height=5.5cm]{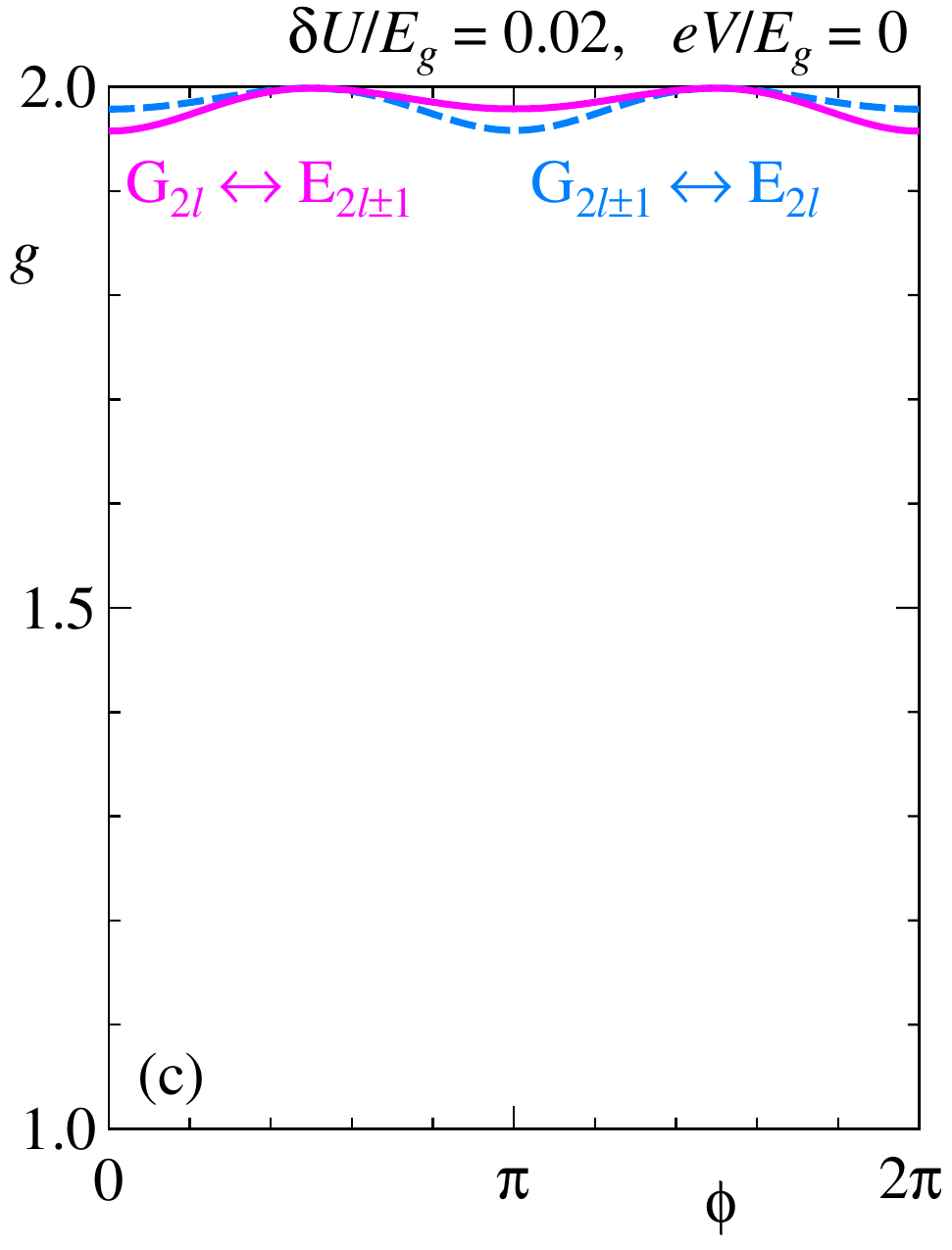}
\end{center}
\end{minipage}
\begin{minipage}{0.5\hsize}
\begin{center}
\hspace{-5mm}
\includegraphics[height=5.5cm]{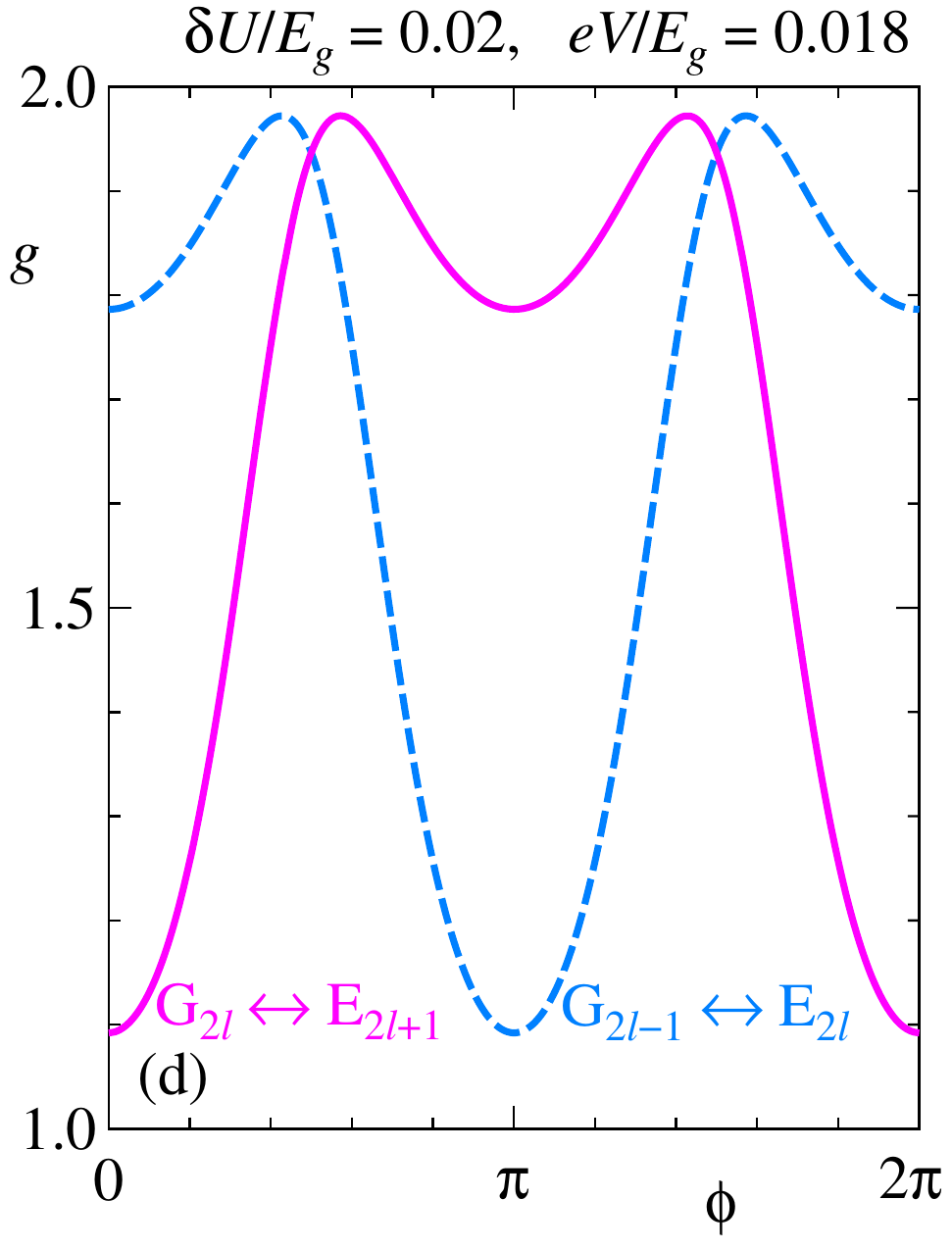}
\end{center}
\end{minipage}
\end{tabular}
\caption{
(Color online) Dimensionless nonlocal conductance $g(V)$ in the presence of
the charging effect with $\delta U/E_{g} =$ (a) $0.0002$, (b) $0.002$,
(c) $0.02$, and (d) $0.02$.
In (a)--(c), solid lines (magenta) represent $g(0)$
in the even ground state ${\rm G}_{2l}$ and dashed lines (blue) represent
$g(0)$ in the odd ground state ${\rm G}_{2l\pm 1}$.
In (d), the solid line (magenta) and the dashed line (blue) represent
$g(V)$ at $eV/E_{g} = 0.018$
under the constrains ${\rm G}_{2l} \leftrightarrow {\rm E}_{2l + 1}$
and ${\rm G}_{2l-1} \leftrightarrow {\rm E}_{2l}$, respectively.
}
\end{figure}
%%%%%%%%%%%%%%%%%%
Figure~3 shows $g(V)$ in the absence of the charging effect
at $eV/E_{g} = 0$ and $0.01$.
The dimensionless nonlocal conductance oscillates
as a function of $\phi$ with a period of $\Phi_{0}/2$.
The amplitude of the oscillation is suppressed
in the zero-bias limit $eV/E_{g} = 0$.~\cite{Sugeta}
This is because, for an incident electron with $E = 0$,
the oscillatory component of the transmission probability to electron
$|t_{\sigma \sigma'}^{e}(E)|^{2}$
is largely cancelled by that of the transmission probability to hole
$|t_{\sigma \sigma'}^{h}(E)|^{2}$.
The values of $\phi$ minimizing $g$ slightly deviate from $0$ and $\pi$,
which reflects the fact that
the spin degeneracy is lifted in the topological superconductor.
For the parameters given above, the true eigenvalue $E_{1}$ normalized by $t$
becomes less than $1.0 \times 10^{-14}$.
We thus set $E_{1}/t = 1.0 \times 10^{-14}$, as described in Sect.~2.
If the terms with $E_{1}$ in the boundary potential cause numerical
instability when $E$ is very close to $E_{1}$,
we need to separate slightly the value of $E$ from $E_{1}$.

Figure~4 shows $g(V)$ in the presence of the charging effect,
where the charging energy is $\delta U/ E_{g} =$
(a) $0.0002$, (b) $0.002$, (c) $0.02$, and (d) 0.02.
In this case, a peak structure due to electron teleportation appears in $g(V)$
near $eV = \delta U$ or $-\delta U$ depending on the constraint.
In (a)--(c), solid lines represent the zero-bias conductance $g(0)$
under the constraint ${\rm G}_{2l} \leftrightarrow {\rm E}_{2l \pm 1}$.
The results for ${\rm G}_{2l} \leftrightarrow {\rm E}_{2l + 1}$
and ${\rm G}_{2l} \leftrightarrow {\rm E}_{2l - 1}$
are not distinguishable at $eV/E_{g} = 0$.
Dashed lines represent $g(0)$ under the constraint
${\rm G}_{2l \pm 1} \leftrightarrow {\rm E}_{2l}$.
Again, the results for ${\rm G}_{2l+1} \leftrightarrow {\rm E}_{2l}$
and ${\rm G}_{2l-1} \leftrightarrow {\rm E}_{2l}$ are not distinguishable.
The overlap between the two results indicates that, for a given $\delta U$,
the oscillation of $g$ due to electron teleportation at $eV = 0$
is determined by only the parity of the ground state.
In (d), the solid and dashed lines represent
$g(V)$ at $eV/E_{g} = 0.018$
under the constraints ${\rm G}_{2l} \leftrightarrow {\rm E}_{2l + 1}$
and ${\rm G}_{2l-1} \leftrightarrow {\rm E}_{2l}$, respectively.
The dimensionless nonlocal conductance oscillates as a function of $\phi$
with a period of $\Phi_{0}$.
The phase shift $\pi$ appears in $g(V)$ between the even and odd ground states.
We observe that the visibility of the phase shift $\pi$
strongly depends on $\delta U$ and $V$.
The unitarity condition,
$\sum_{\sigma}(|r_{\sigma \sigma'}^{e}|^{2}+|t_{\sigma \sigma'}^{e}|^{2}) = 1$,
is checked in each calculation.

\section{Summary and Discussion}

We considered the conductance of an interferometer containing
a topological superconductor that accommodates a pair of Majorana zero modes.
The conductance is significantly affected by electron teleportation
owing to the nonlocal state consisting of the Majorana zero modes
when the number of electrons $\mathcal N$ in the superconductor
is constrained by a charging effect.
We presented a boundary potential method for calculating the conductance
under a given constraint on $\mathcal N$ with minimal computational effort.
To demonstrate its usefulness, we obtained the nonlocal conductance
under several different constraints.
A slight drawback of this method is
that it cannot incorporate inelastic scattering processes.
The boundary potential method is useful for simulating the conductance
of the interferometer incorporating the details of the system,
such as the geometry of the interferometer,
the spatial variation of potential in the superconductor,
and the coupling strength between the superconductor and the lead.
The results of such simulations will be reported in a separate publication.

Finally, we show that the phase shift $\pi$ is encoded in
$\bigl[V_{1 M}^{e\sigma, e\sigma'}(E)\bigr]_{\pm}^{{\mathcal E}/{\mathcal O}}$
and
$\bigl[V_{M 1}^{e\sigma, e\sigma'}(E)\bigr]_{\pm}^{{\mathcal E}/{\mathcal O}}$,
which describe the electron teleportation.
We focus on the contribution from the nonlocal state to
$\bigl[V_{1 M}^{e\sigma, e\sigma'}(E)\bigr]_{\pm}^{{\mathcal E}/{\mathcal O}}$
in the case $E = 0$ with $E_{1} = 0$ for simplicity
and refer to the resulting contribution as
$\bigl[\tilde{V}_{1 M}^{e\sigma, e\sigma'}
\bigr]_{\pm}^{{\mathcal E}/{\mathcal O}}$.
From Eqs.~(\ref{eq:V_1M_+E}), (\ref{eq:V_1M_-E}), (\ref{eq:V_1M_+O}),
and (\ref{eq:V_1M_-O}), we find
\begin{align}
  \bigl[\tilde{V}_{1 M}^{e\sigma, e\sigma'}\bigr]_{+}^{\mathcal E}
 & = -e^{-i\phi}\frac{t_{1}t_{2}}{\delta U}
      u_{1\sigma}(1)u_{1\sigma'}(N) ,
    \\
  \bigl[\tilde{V}_{1 M}^{e\sigma, e\sigma'}\bigr]_{-}^{\mathcal E}
 & = e^{-i\phi}\frac{t_{1}t_{2}}{\delta U}
      v_{1\sigma}(1)v_{1\sigma'}(N)
\end{align}
in the even ground state ${\rm G}_{2l}$, and
\begin{align}
  \bigl[\tilde{V}_{1 M}^{e\sigma, e\sigma'}\bigr]_{+}^{\mathcal O}
 & = -e^{-i\phi}\frac{t_{1}t_{2}}{\delta U}
      v_{1\sigma}(1)v_{1\sigma'}(N) ,
    \\
  \bigl[\tilde{V}_{1 M}^{e\sigma, e\sigma'}\bigr]_{-}^{\mathcal O}
 & =  e^{-i\phi}\frac{t_{1}t_{2}}{\delta U}
      u_{1\sigma}(1)u_{1\sigma'}(N) 
\end{align}
in the odd ground state ${\rm G}_{2l\pm 1}$.
The particle--hole symmetry with $E_{1} = 0$ ensures
$u_{1\sigma}(1)u_{1\sigma'}(N) = -v_{1\sigma}(1)v_{1\sigma'}(N)$.
We thus find
\begin{align}
   \bigl[\tilde{V}_{1 M}^{e\sigma, e\sigma'}\bigr]_{+}^{\mathcal E}
 = \bigl[\tilde{V}_{1 M}^{e\sigma, e\sigma'}\bigr]_{-}^{\mathcal E}
 = -\bigl[\tilde{V}_{1 M}^{e\sigma, e\sigma'}\bigr]_{+}^{\mathcal O}
 = -\bigl[\tilde{V}_{1 M}^{e\sigma, e\sigma'}\bigr]_{-}^{\mathcal O} .
\end{align}
This explains the phase shift of $\pi$ that is induced by the change in
the parity in the ground state.
Its mechanism is similar to that of the parity-change-driven $0$--$\pi$
transition in Josephson junctions.~\cite{Spivak,Bauernschmitt,Shimizu,vanDam}

\appendix

\section{Majorana Operators}

We show that Majorana operators can be constructed using
$|\varphi_{+}\rangle$ and $|\varphi_{-}\rangle$ (see Sect.~2)
in the case of a sufficiently large $N$.
These eigenvectors satisfy
$\Xi|\varphi_{\pm}\rangle = \pm |\varphi_{\pm}\rangle$,
and their elements are real.
According to Eq.~(\ref{eq:def-d}), we define $\psi_{1}$ and
$\psi_{2}$ as
\begin{align}
    \label{eq:def-psi_1}
  \psi_{1} & = \sqrt{2} C^{\dagger}|\varphi_{+}\rangle ,
  \\
    \label{eq:def-psi_2}
  \psi_{2} & = \sqrt{2} C^{\dagger}(-i)|\varphi_{-}\rangle .
\end{align}
Since $\Xi|\varphi_{+}\rangle = |\varphi_{+}\rangle$,
the electron and hole components of $|\varphi_{+}\rangle$ have
the same amplitude and sign at each site;
thus, we find $\psi_{1}=\psi_{1}^{\dagger}$.
Since $\Xi|\varphi_{-}\rangle = -|\varphi_{-}\rangle$,
the electron and hole components of $|\varphi_{-}\rangle$ also have
the same amplitude at each site, but their signs are reversed.
The factor $-i$ is added to the definition of $\psi_{2}$
so that $\psi_{2}=\psi_{2}^{\dagger}$.
The factor $\sqrt{2}$ ensures $\psi_{1}^{2}=\psi_{2}^{2} = 1$.
From $\langle\chi_{\pm}|\chi_{\mp}\rangle = 0$,
we find that these operators anticommute with each other.
The above results show that $\psi_{1}$ and $\psi_{2}$ are
a pair of Majorana operators satisfying
\begin{align}
  \psi_{i}\psi_{j}+\psi_{j}\psi_{i} = 2\delta_{i,j}
\end{align}
for $i, j$ = $1, 2$.
From $\langle\chi_{\pm}|\varphi_{n}\rangle = 0$
for $n=\pm2, \pm3, \dots, \pm2N$,
we find that $\psi_{i}$ anticommutes with the bogolon operators
$d_{n}$ and $d_{n}^{\dagger}$ for $n=2, 3, \dots, 2N$.

The creation and annihilation operators for the nonlocal state are
expressed as
\begin{align}
  d_{1}^{\dagger}
   & = \frac{1}{2}\left(\psi_{1}+i\psi_{2}\right) ,
  \\
  d_{1}
   & = \frac{1}{2}\left(\psi_{1}-i\psi_{2}\right) .
\end{align}
Using Eqs.~(\ref{eq:def-psi_1}) and (\ref{eq:def-psi_2}) with
Eqs.~(\ref{eq:def-varphi_1}) and (\ref{eq:def-tildevarphi_1}),
we can rewrite the creation and annihilation operators as
\begin{align}
  d_{1}^{\dagger} = C^{\dagger}|\varphi_{1}\rangle, \hspace{5mm}
  d_{1} = C^{\dagger}|\varphi_{-1}\rangle ,
\end{align}
which are consistent with the definitions of $d_{1}$ and $d_{1}^{\dagger}$.

\section{Derivation of the Boundary Potential}

Carrying out the integrations over the Grassmann variables
in Eq.~(\ref{eq:def-S_B}), we find
\begin{align}
   S_{\rm B}
 & = - \sum_{\omega}\sum_{n=1}^{2N}\frac{1}{-i\omega + E_{n}}
       \sum_{\sigma}\sum_{\sigma'}
           \nonumber \\
 & \hspace{-3mm}\times
   \biggl[ -t_{1}
            \left( u_{n\sigma}(1)\bar{f}_{1\sigma}(\omega)
                  -v_{n\sigma}(1)f_{1\sigma}(-\omega)
            \right)
       \nonumber \\
 & \hspace{-3mm}
           -t_{2}
            \left( e^{i\phi}u_{n\sigma}(N)\bar{f}_{M\sigma}(\omega)
                  -e^{-i\phi}v_{n\sigma}(N)f_{M\sigma}(-\omega)
            \right)
   \biggr]
       \nonumber \\
 & \hspace{-3mm} \times
   \biggl[ -t_{1}
            \left( u_{n\sigma'}(1)f_{1\sigma'}(\omega)
                  -v_{n\sigma'}(1)\bar{f}_{1\sigma'}(-\omega)
            \right)
       \nonumber \\
 & \hspace{-3mm}
           -t_{2}
            \left( e^{-i\phi}u_{n\sigma'}(N)f_{M\sigma'}(\omega)
                  -e^{i\phi}v_{n\sigma'}(N)\bar{f}_{M\sigma'}(-\omega)
            \right)
   \biggr] .
\end{align}
After arranging the terms in the above expression
in the form of Eq.~(\ref{eq:def-S-B-Mat}), we find
\begin{align}
  V_{1 1}^{e\sigma, e\sigma'}(\omega)
 & = -t_{1}^{2}\sum_{n=1}^{2N}
%      \nonumber \\
% & \hspace{-12mm}
%      \times
      \left( \frac{u_{n\sigma}(1)u_{n\sigma'}(1)}{-i\omega+E_{n}}
           - \frac{v_{n\sigma}(1)v_{n\sigma'}(1)}{i\omega+E_{n}}
      \right) ,
       \\
  V_{1 1}^{h\sigma, h\sigma'}(\omega)
 & = -t_{1}^{2}\sum_{n=1}^{2N}
%      \nonumber \\
% & \hspace{-12mm}
%      \times
      \left( \frac{v_{n\sigma}(1)v_{n\sigma'}(1)}{-i\omega+E_{n}}
           - \frac{u_{n\sigma}(1)u_{n\sigma'}(1)}{i\omega+E_{n}}
      \right) ,
       \\
  V_{1 1}^{e\sigma, h\sigma'}(\omega)
 & = t_{1}^{2}\sum_{n=1}^{2N}
%      \nonumber \\
% & \hspace{-12mm}
%      \times
      \left( \frac{u_{n\sigma}(1)v_{n\sigma'}(1)}{-i\omega+E_{n}}
           - \frac{v_{n\sigma}(1)u_{n\sigma'}(1)}{i\omega+E_{n}}
      \right) ,
       \\
  V_{1 1}^{h\sigma, e\sigma'}(\omega)
 & = t_{1}^{2}\sum_{n=1}^{2N}
%      \nonumber \\
% & \hspace{-12mm}
%      \times
      \left( \frac{v_{n\sigma}(1)u_{n\sigma'}(1)}{-i\omega+E_{n}}
           - \frac{u_{n\sigma}(1)v_{n\sigma'}(1)}{i\omega+E_{n}}
      \right) ,
       \\
  V_{M M}^{e\sigma, e\sigma'}(\omega)
 & = -t_{2}^{2}\sum_{n=1}^{2N}
      \nonumber \\
 & \hspace{-12mm}
      \times
      \left( \frac{u_{n\sigma}(N)u_{n\sigma'}(N)}{-i\omega+E_{n}}
           - \frac{v_{n\sigma}(N)v_{n\sigma'}(N)}{i\omega+E_{n}}
      \right) ,
       \\
  V_{M M}^{h\sigma, h\sigma'}(\omega)
 & = -t_{2}^{2}\sum_{n=1}^{2N}
      \nonumber \\
 & \hspace{-12mm}
      \times
      \left( \frac{v_{n\sigma}(N)v_{n\sigma'}(N)}{-i\omega+E_{n}}
           - \frac{u_{n\sigma}(N)u_{n\sigma'}(N)}{i\omega+E_{n}}
      \right) ,
       \\
  V_{M M}^{e\sigma, h\sigma'}(\omega)
 & = e^{i2\phi}t_{2}^{2}\sum_{n=1}^{2N}
      \nonumber \\
 & \hspace{-12mm}
      \times
      \left( \frac{u_{n\sigma}(N)v_{n\sigma'}(N)}{-i\omega+E_{n}}
           - \frac{v_{n\sigma}(N)u_{n\sigma'}(N)}{i\omega+E_{n}}
      \right) ,
       \\
  V_{M M}^{h\sigma, e\sigma'}(\omega)
 & = e^{-i2\phi}t_{2}^{2}\sum_{n=1}^{2N}
      \nonumber \\
 & \hspace{-12mm}
      \times
      \left( \frac{v_{n\sigma}(N)u_{n\sigma'}(N)}{-i\omega+E_{n}}
           - \frac{u_{n\sigma}(N)v_{n\sigma'}(N)}{i\omega+E_{n}}
      \right) ,
       \\
  V_{1 M}^{e\sigma, e\sigma'}(\omega)
 & = -e^{-i\phi}t_{1}t_{2}\sum_{n=1}^{2N}
      \nonumber \\
 & \hspace{-12mm}
      \times
      \left( \frac{u_{n\sigma}(1)u_{n\sigma'}(N)}{-i\omega+E_{n}}
           - \frac{v_{n\sigma}(1)v_{n\sigma'}(N)}{i\omega+E_{n}}
      \right) ,
       \\
  V_{1 M}^{h\sigma, h\sigma'}(\omega)
 & = -e^{i\phi}t_{1}t_{2}\sum_{n=1}^{2N}
      \nonumber \\
 & \hspace{-12mm}
      \times
      \left( \frac{v_{n\sigma}(1)v_{n\sigma'}(N)}{-i\omega+E_{n}}
           - \frac{u_{n\sigma}(1)u_{n\sigma'}(N)}{i\omega+E_{n}}
      \right) ,
       \\
  V_{1 M}^{e\sigma, h\sigma'}(\omega)
 & = e^{i\phi}t_{1}t_{2}\sum_{n=1}^{2N}
      \nonumber \\
 & \hspace{-12mm}
      \times
      \left( \frac{u_{n\sigma}(1)v_{n\sigma'}(N)}{-i\omega+E_{n}}
           - \frac{v_{n\sigma}(1)u_{n\sigma'}(N)}{i\omega+E_{n}}
      \right) ,
       \\
  V_{1 M}^{h\sigma, e\sigma'}(\omega)
 & = e^{-i\phi}t_{1}t_{2}\sum_{n=1}^{2N}
      \nonumber \\
 & \hspace{-12mm}
      \times
      \left( \frac{v_{n\sigma}(1)u_{n\sigma'}(N)}{-i\omega+E_{n}}
           - \frac{u_{n\sigma}(1)v_{n\sigma'}(N)}{i\omega+E_{n}}
      \right) ,
       \\
  V_{M 1}^{e\sigma, e\sigma'}(\omega)
 & = -e^{i\phi}t_{1}t_{2}\sum_{n=1}^{2N}
      \nonumber \\
 & \hspace{-12mm}
      \times
      \left( \frac{u_{n\sigma}(N)u_{n\sigma'}(1)}{-i\omega+E_{n}}
           - \frac{v_{n\sigma}(N)v_{n\sigma'}(1)}{i\omega+E_{n}}
      \right) ,
       \\
  V_{M 1}^{h\sigma, h\sigma'}(\omega)
 & = -e^{-i\phi}t_{1}t_{2}\sum_{n=1}^{2N}
      \nonumber \\
 & \hspace{-12mm}
      \times
      \left( \frac{v_{n\sigma}(N)v_{n\sigma'}(1)}{-i\omega+E_{n}}
           - \frac{u_{n\sigma}(N)u_{n\sigma'}(1)}{i\omega+E_{n}}
      \right) ,
       \\
  V_{M 1}^{e\sigma, h\sigma'}(\omega)
 & = e^{i\phi}t_{1}t_{2}\sum_{n=1}^{2N}
      \nonumber \\
 & \hspace{-12mm}
      \times
      \left( \frac{u_{n\sigma}(N)v_{n\sigma'}(1)}{-i\omega+E_{n}}
           - \frac{v_{n\sigma}(N)u_{n\sigma'}(1)}{i\omega+E_{n}}
      \right) ,
       \\
  V_{M 1}^{h\sigma, e\sigma'}(\omega)
 & = e^{-i\phi}t_{1}t_{2}\sum_{n=1}^{2N}
      \nonumber \\
 & \hspace{-12mm}
      \times
      \left( \frac{v_{n\sigma}(N)u_{n\sigma'}(1)}{-i\omega+E_{n}}
           - \frac{u_{n\sigma}(N)v_{n\sigma'}(1)}{i\omega+E_{n}}
      \right) .
\end{align}

\end{document}